\documentclass[10pt,a4paper]{article}

\usepackage[T1]{fontenc}
\usepackage[utf8]{inputenc}
\usepackage{authblk} 
\usepackage{amssymb}
\usepackage{amsmath}
\usepackage{amsthm}
\usepackage{mathtools} 
\usepackage{latexsym}
\usepackage{graphicx}
\usepackage{dutchcal}
\usepackage{tcolorbox}
\usepackage{wrapfig}
\usepackage{tikz}
\usepackage{color}
\usepackage[font={footnotesize}]{caption}
\usepackage[font={footnotesize}]{subcaption}

\usepackage{hyperref}
\usepackage{algorithmicx}
\usepackage{lipsum}

\usepackage{csquotes}
\usepackage{url}
\usepackage{epsf}

\usepackage{float}

\usepackage{widetext}
\usepackage{cuted,multicol} 

\usepackage{arydshln}

\usepackage{sectsty}
\sectionfont{\centering}

\theoremstyle{definition}
\newtheorem{thm}{Theorem}[section]
\newtheorem{lem}[thm]{Lemma}
\newtheorem{prop}[thm]{Proposition}

\newenvironment{pf}{{\noindent\sc Proof. }}{\qed}

\theoremstyle{definition}
\newtheorem*{defn}{Definition}

\theoremstyle{definition}

\newtheorem{asmpn}{Assumption}[section]

\providecommand{\keywords}[1]{\textbf{\textit{Index terms---}} #1}

\newcommand{\reals}{{\mathbb R}}

\newcommand{\E}{{\mathbb E}}

\newcommand{\x}{{\mathbf x}}
\newcommand{\xc}{\mathbf{x}^{c_i}}
\newcommand{\xp}{\mathbf{x}^{p_j}}

\newcommand{\p}{{\mathbf p}}

\renewcommand{\u}{{\mathbf u}}
\renewcommand{\b}{{\mathbf b}}
\newcommand{\nos}{{\mathbf n}}
\newcommand{\bc}{\mathbf{b}^{c_i}}
\newcommand{\bp}{\mathbf{b}^{p_j}}

\DeclareMathOperator*{\argmax}{arg\,max}
\definecolor{awesome}{rgb}{1.0, 0.13, 0.32}
\definecolor{ao}{rgb}{0.0, 0.0, 1.0}
\definecolor{cadmiumgreen}{rgb}{0.0, 0.42, 0.24}
\definecolor{darkpastelpurple}{rgb}{0.59, 0.44, 0.84}
\definecolor{darkorchid}{rgb}{0.6, 0.2, 0.8}
\definecolor{lightblue}{rgb}{0.68, 0.85, 0.9}
\definecolor{lavenderblue}{rgb}{0.8, 0.8, 1.0}
\definecolor{amethyst}{rgb}{0.6, 0.2, 0.8}
\definecolor{ao(english)}{rgb}{0.0, 0.5, 0.0}
\definecolor{alizarin}{rgb}{0.82, 0.1, 0.26}
\definecolor{amber(sae/ece)}{rgb}{1.0, 0.49, 0.0}
\definecolor{cadetgrey}{rgb}{0.57, 0.64, 0.69}
\definecolor{internationalkleinblue}{rgb}{0.0, 0.18, 0.65}
\definecolor{electricultramarine}{rgb}{0.25, 0.0, 1.0}
\definecolor{frenchblue}{rgb}{0.0, 0.45, 0.73}

\title{On the Volatility of Optimal Control Policies \\ and the Capacity of a Class of Linear Quadratic Regulators\footnote{\color{red}Under review. Please do not distribute.}}

\author[1]{Avi Mohan\thanks{avinashmohan@campus.technion.ac.il ({\color{blue}Corresponding author})}}
\author[1]{Shie Mannor\thanks{shie@ee.technion.ac.il}}
\author[2]{Arman C. Kizilkale\thanks{akizilkale@samasource.org}}
\affil[1]{Faculty of Electrical Engineering, Technion, Israel Institute of Technology}
\affil[2]{Samasource, Montr\'eal, Canada}

\date{}
\begin{document}
\maketitle





\begin{abstract}

  It is well known that, despite being mathematically sound, control laws that display high temporal variability are undesirable from an engineering perspective. The effects of volatility in the control input manifest in myriad ways depending on the application and are invariably deleterious to the controlled system. In this article we are concerned with the temporal volatility of the regulator’s control process in discrete time Linear Quadratic Regulators (LQRs). Our investigation in this paper unearths a surprising and very fundamental connection between the cost functional which the LQR is tasked with minimizing and the temporal variations of its control laws. 
  
  We first show that optimally controlling the system always implies high levels of control volatility, i.e., it is impossible to reduce volatility in the optimal control process without sacrificing cost. We also show that, akin to communication systems, every LQR has a \emph{\color{blue}Capacity Region} associated with it, that dictates and quantifies how much cost is achievable at a given level of control volatility. This additionally establishes the fact that no admissible control policy can simultaneously achieve low volatility and low cost. 
   
   We then employ this analysis to explain an important phenomenon observed in deregulated electricity markets. Spot prices in such markets in the United States and Europe have historically displayed high volatility, leading to loss of revenue in high electricity costs, weakening economic growth and lost business from long blackouts. With rapidly increasing supply of intermittent solar and wind-based power, conditions in the foreseeable future are only expected to deteriorate. Geographic price variations, while capable of being quite dramatic themselves, are a natural outcome of Locational Marginal Pricing (LMP) that takes finite transmission capacity and congestion into consideration, to produce localized electricity prices. \emph{Temporal} price volatility (considerable fluctuation over periods as short as an hour in particular), however, is harder to explain.
  
  In an effort to unearth the cause of these price fluctuations, we analyze the electricity market when composed of suppliers and consumers who are (i) strategic, \emph{\color{blue}price-anticipating} participants and (ii) passive, \emph{\color{blue}price-taking} participants. In the former case, participants submit bids, in the form of price-quantity graphs, to the market operator (RTO/ISO), which then computes and advertizes the spot price of electricity. Here, we show that in {\em any} such market, at the Nash-Walras equilibrium, improving social welfare must necessarily be traded off with volatility in prices, i.e., it is impossible to reduce volatility in the price of electricity without sacrificing social welfare. In the latter case, we model the electricity market as a discrete time controlled stochastic process, and arrive at the same conclusion. Going a step further, in the context of renewable power sources, our investigation uncovers an intriguing phenomenon we term the \emph{\color{blue}volatility cliff}, which suggests that with increasing penetration of intermittent renewable production, price volatility could increase to unacceptable levels, prompting the need for a complete restructuring of  existing electricity markets.
  
\end{abstract}


\tableofcontents

\section{Introduction}\label{sec:Introduction}
The Linear Quadratic Regulator (LQR) is one of the cornerstones of optimal control theory \cite{kumar-varaiya15stochastic-systems-estimation-adaptive-control,bertsekas95dynamic-programming-optimal-control-volume1}. It is concerned with operating a \emph{linear} dynamical system in such a way as to minimize a cost functional that is \emph{quadratic} in the system state and control input. Over the years, the LQR has been used as an analytical model in fields as varied as aeronautical engineering \cite{ben-asher10optimal-control-aerospace}, chemical engineering \cite{binder-etal01introduction-model-optimization-chemical-processes} and economic theory \cite{carraro-sartore12developments-control-economic-analysis,chow75analysis-control-dynamic-economic-systems}. The dynamical system in the LRQ model can evolve in either discrete time or continuous time and this problem has been subjected to extensive investigation in both \cite{kumar-varaiya15stochastic-systems-estimation-adaptive-control}. 

In this article we are concerned with the \emph{temporal volatility} of the regulator's control process in discrete time LQRs. It is well known that, despite being mathematically sound, highly volatile control laws are undesirable from an engineering perspective \cite{gao04discrete-time-optimal-closed-form}. The high frequency switching these laws entail are difficult to implement and tend to excite undesired second-order dynamics \cite{bohez-etal19value-constrained-model-free-continuous-control}. Indeed frequent and abrupt switching in bang-bang control for example, invariably leads to a phenomenon called \enquote{control signal chattering} \cite{gao-shaohua04properties-new-form-discrete-optimal}, that tends to rapidly wear out actuators. The deleterious effects of wide temporal swings in the control input manifest in myriad ways depending on the application. In fact, in the sequel, we will provide multiple examples of this phenomenon from the field of economic theory. Our investigation in this paper unearths a surprising and very fundamental connection between the aforementioned cost functional and the temporal variations of the control law. We initially show that optimally controlling the system always implies high levels of control volatility. Thereafter, we extend this result to show that no \emph{admissible}\footnote{See Sec.~\ref{sec:lqrModelPreliminaries} for the definition of admissibility.} control can simultaneously achieve low volatility and low cost. We then use these results to help explain the phenomenon of price volatility in deregulated electricity markets.

\subsection{Price Volatility in Deregulated Markets}
The deregulated electricity market in the United States is managed by two types of profit-neutral entities called Independent System Operators (ISOs) and Regional Transmission Organizations (RTOs). These entities are primarily responsible for the reliability of the power grid, resource planning i.e., keeping the power grid balanced between generation (supply) and load (demand), and determining electricity prices which, in turn, heavily influence supply, demand and future investment in the power sector. In fact, there is general consensus that setting real-time prices that reflect current operating conditions has multiple positive effects on the market such as the potential to reduce supplier ancillary cost and improving system efficiency \cite{department-energy06benefits-demand-response}. Consequently, the price of electricity, which we shall also refer to as the price \emph{signal,} has assumed critical importance in wholesale power markets \cite{hogan10demand-response-compensation-benefits}.

\begin{figure}[tbh]
\centering
\includegraphics[height=6.00cm, width=10.00cm]{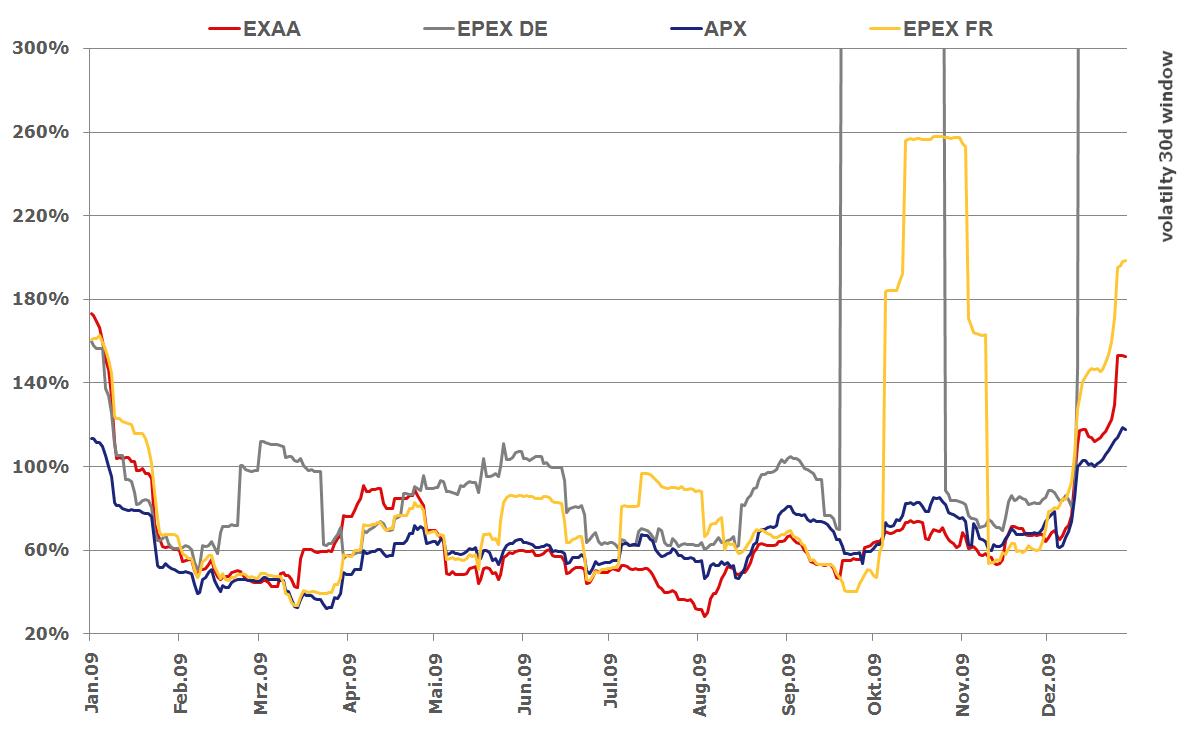}
\caption{Market volatility of electricity prices in European power markets in 2009. The figure compares prices in four electricity exchanges namely, the Austrian energy exchange ({\color{red}red}), European power exchange (EPEX) Germany ({\color{gray}gray}), EPEX France ({\color{orange}yellow}) and the APX group ({\color{internationalkleinblue}blue}). Of particular interest is the EPEX Germany price curve that fluctuates much more than $300\%$ in Oct 2009 alone. Source: Wikimedia Commons \cite{exaa10volatility-price-european-markets}.}
\label{fig:priceVolatilityEuropeanMarkets}
\end{figure}

However, in the years following the deregulation of the electricity market, volatile electricity prices have turned into an epidemic showing jumps of up to $300\%$ over periods as short as one hour \cite{solomon15why-electricity-prices-volatile}, and this phenomenon has been highlighted by researchers as a \emph{feature} of the market \cite{cho-meyn10efficiency-marginal-cost-dynamic-competitive-markets-friction, benini-eta02day-ahead-volatility-analysis,robinson05math-model-explains-prices-electricity,kizilkale-etal10regulation-efficiency-markets-friction} rather than due to isolated cases of exercise of market power such as that involving the Enron Corporation \cite{fox03enron-rise-fall}. European markets, as seen in Fig.~\ref{fig:priceVolatilityEuropeanMarkets}, have been experiencing similar fluctuations in prices. Additionally, accounting for transmission constraints across the grid and demand uncertainty localization has resulted in sharp geographical price variations. This mechanism, called \emph{Locational Marginal Pricing} (LMP) can result in dramatic disparities in electricity prices such as in June 2010, when the price across a single RTO (the Pennsylvania, Jersey, Maryland Power Pool or PJM) varied by $635.65\%$ \cite{shmuel11primer-competitive-electricity-smart}. 

\begin{figure}[tbh]
\hspace{-0.5cm}%
\centering
\includegraphics[height=6.50cm, width=9.50cm]{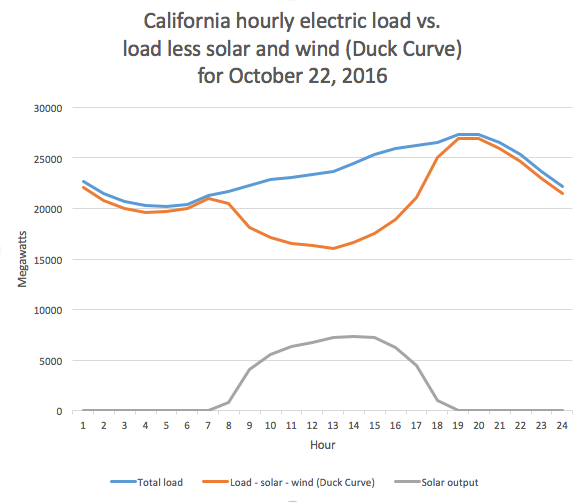}
\caption{The infamous California ISO \enquote{Duck curve.} The {\color{amber(sae/ece)}orange} plot represents the aggregate power generated over a period of 24 hours by renewable sources (wind and solar), while the {\color{blue}blue} plot represents the total demand on the system. The {\bf\color{gray}gray} curve at the bottom shows the contribution from solar power sources.  As renewable sources proliferate, this effect is expected to only worsen, as evidenced by states such as Hawaii, where a more pronounced version of this phenomenon, called the \enquote{Nessie curve} has been observed \cite{stjohn14hawaii-solar-nessie}. Source: Wikimedia Commons \cite{reinhold-etal16caiso-duck-curve-diagram-wikimedia}.}
\label{fig:duckCurveInCAISO}
\end{figure}

Widespread adoption of \enquote{green} renewable energy sources such as windmills and photovoltaic (PV) cells is beginning to exacerbate this problem and has even led to long periods of \emph{negative} electricity prices \cite{goetz17negative-prices-spot-market, aleasoft19wind-energy-negative-prices-germany}, with suppliers paying to keep their generation plants running. Fig.~\ref{fig:duckCurveInCAISO} illustrates the infamous \enquote{duck curve} that regularly occurs in locations such as within the California ISO, where a substantial amount of solar electric capacity has been installed \cite{denholm-etal15overgeneration-solar-california-duck-chart}. Following periods of high solar generation during daytime, power suppliers need to rapidly increase their output using conventional sources (such as natural gas) around sunset to compensate for the sudden fall in solar power levels. This precipitous increase in demand only adds to the fluctuations in electricity prices. While improvements in grid energy storage technologies such as battery storage, pumped storage hydroelectricity and compressed air look promising, they suffer from multiple limitations such as prohibitively high investment costs, efficiency, need for appropriate geography, environmental impacts etc., and do not appear to be capable of solving the volatility problem at least in the near future \cite{wiki19pumped-hydroelectricity,yang-jackson11opportunities-barriers-pumped-hydro-storage-us,wiki19compressed-air-storage}. It is notable that, since the late 1990s, all these price variations have resulted in an estimated $\$45$ billion in lost business from long blackouts, higher electricity costs and weakening economic growth in California alone \cite{robinson05math-model-explains-prices-electricity}. 

\subsection{Related Work}\label{sec:relatedWork}
Analysis of this volatility phenomenon has received some attention in recent times. The work in  \cite{cho-meyn10efficiency-marginal-cost-dynamic-competitive-markets-friction,kizilkale-etal10regulation-efficiency-markets-friction} considered continuous time markets with friction, i.e., bounded rate of change of generation and/or demand, and showed that the optimal price signal, while not harming (a suitably defined notion of) social utility\footnote{We use the terms \enquote{social welfare} and \enquote{social utility} interchangeably.}, does tend to fluctuate widely. The former article accounts only for supply-side friction and demonstrates using a stylized model, that the utility optimizing price tends to show extreme fluctuations in time. The latter expands this model to include demand-side friction in a market with Brownian activation, and demonstrates a similar result. \cite{kizilkale-etal10regulation-efficiency-markets-friction} also shows a tradeoff between social utility and price volatility for this optimal market price process. 

Two points are of interest here. Firstly, RTOs and ISOs run electricity markets at a minimum temporal granularity of about 5-15 minutes \cite{dahleh-mitter12volatility-power-grids-real-time,yang-ozdaglar16reducing-electricity-price-volatility-storage}. This means that generation, demand and prices, even in real-time markets, are reevaluated only every 5-15 minutes\footnote{In fact, this is true in most regions of the United States such as New England, the PJM and California \cite{dahleh-mitter12volatility-power-grids-real-time}}. 
These markets, therefore, are not modeled well with continuous time dynamics and require a discrete time model for analysis. But this immediately renders the proof techniques in the aforementioned work unusable and we, in the current article, develop {entirely new} analytical machinery specifically for markets that operate in discrete time. Secondly, neither of these articles comments about the volatility behavior of other pricing policies. In the present article, we go further and show that volatility and social utility necessarily have to be traded off regardless of what pricing policy is used. We also completely characterize the space of (volatility,utility) tuples achievable in any given market. We hope that, akin to what the Shannon capacity region \cite{cover-thomas12elements-information-theory} did for communication systems, this result will help guide the development of pricing policies for such markets in general.

The electricity market is modeled as a \emph{deterministic} discrete time dynamical system without supply-side friction in  \cite{dahleh-mitter12volatility-power-grids-real-time}. The article analyzes social cost minimization under market clearing constraints, defines a Lyapunov stability notion for the price process and derives conditions under which stability can be guaranteed. It also defines two notions of volatility: aggregate and time averaged, for price, demand and generation processes, and shows that when market dynamics satisfy certain smoothness assumptions, bounded volatility can be guaranteed.
\cite{wei-etal14competitive-equilibrium-electricity-price-fluctuation} considers a deterministic finite time horizon (market closes after $T$ time slots) market with a single supplier and multiple consumers. It defines a bounded quasilinear model of consumer utility (i.e., Consumer~$i$ consumes at most $d_i$ units over $T$ slots), and derives a pricing mechanism to maximize social utility under an appropriately defined price fluctuation penalty. \cite{wei-etal14competitive-equilibrium-electricity-price-fluctuation} also provides a subgradient descent-based algorithm for implementing utility maximization in a decentralized manner. More recently, \cite{singh-etal18decentralized-control-dynamic-prices-independent-system-operator} has adopted a game theoretic approach to solve this problem in the absence of the knowledge of consumer preferences (utility functions).

Turning to geographic price volatility, \cite{yang-ozdaglar16reducing-electricity-price-volatility-storage} adopts a stochastic control-based approach to analyze the impact of grid storage in mitigating sharp fluctuations in Locational Marginal Pricing (LMP). 
Increase in price variability due to widespread adoption of renewable sources could be observed as far back as 2011, when empirical studies analyzing spot-price variance using data from the Electricity Reliability Council of Texas (ERCOT) concluded that increased adoption of wind-based power sources tends to decrease spot prices on an average, but increases their temporal variability \cite{woo-etal11impact-wind-generation-electricity-price-variance}, i.e., prices tend to fluctuate more widely and faster. 
More recently, similar analysis of data from the California ISO also draws similar conclusions \cite{joskow19challenges-wholesale-intermittent-renewable-at-scale}.


\subsection{Our contributions and organization}\label{sec:ourContributionsAndOrganization}
We now provide a road-map of the rest of the article together with a summary of our contributions. 
\begin{itemize}
    \item We begin with the Linear Quadratic Regulator model in Sec.~\ref{sec:lqrModelPreliminaries}. We define notions of \enquote{admissibility} of a control policy, \enquote{regulator efficiency} and \enquote{control volatility} that we analyze in depth in future sections of the paper.
    
    \item We then study a tradeoff between the aforementioned regulator efficiency and volatility of the control law in Sec.~\ref{sec:regulatorEffVersusControlVolatilityinLQR}. We prove that penalizing volatility in the optimal (in the sense of minimizing cost) control process automatically penalizes efficiency. We then provide a complete characterization of the set of all achievable (volatility, efficiency) tuples for any given LQR (Sec.~\ref{sec:CapacityRegionNoFreeLunchTheorem}) which we call the \emph{\color{blue}capacity region} associated with the LQR. This automatically gives rise to a \enquote{No free lunch} theorem that states that no admissible control policy can simultaneously achieve low volatility and low cost.
    
    \item We then use the above results to explain the dramatic levels of price fluctuations in real-time electricity markets in the United States and Europe (Sec.~\ref{sec:marketsWithPriceAnticipatingParticipants} and Sec.~\ref{sec:applicationPriceTakingMarketAnalysis}). To begin with, we consider an electricity market with strategic, price-anticipating suppliers and consumers. We model the market as an infinite sequence of auctions between buyers and sellers who submit bids for electric power to the market operator (i.e., RTO/ISO) in the form of price-quantity (PQ) graphs. The operator in turn decides and advertizes the price of electricity as a function of these bids. We derive a Nash equilibrium strategy profile for this dynamic game and show (i) that the equilibrium is Stationary Markov Perfect, and (ii) using the analytical techniques developed in the previous sections, that efficient market operation entails volatile electricity prices. 
    
    \item In anticipation of concerns that the strategic or competitive behavior of market participants might be at the root of this phenomenon, we next show that even with passive prosumers, this market behavior cannot be avoided. In Sec.~\ref{sec:applicationPriceTakingMarketAnalysis}, we model the market as composed of passive, price-taking participants, i.e., the market is now modeled as a discrete-time controlled stochastic process. Here, once again, our analysis shows that the aforementioned efficiency-volatility tradeoff is unavoidable.  Next, we analyze the effect of \emph{decarbonization}, i.e., drastically increasing dependence on intermittent solar and wind generation, on price volatility (Sec.~\ref{sec:renewablesAndTheVolatilityCliff}). Our study suggests the occurrence of a \emph{volatility cliff} phenomenon which predicts that, with increasing penetration of intermittent renewable production, price fluctuation could increase to unacceptable levels. Echoing the assertions made in \cite{joskow19challenges-wholesale-intermittent-renewable-at-scale} for renewed interest in electricity market reform,  we hope that this result aids in convincing the reader of the need for a complete revamp of the existing electricity market structure to handle large-scale adoption of intermittent power sources.
    
    \item We then present simulations to help illustrate our results in Sec.~\ref{sec:simulationResults}. Finally, we provide concluding remarks and discuss possible directions for future work (Sec.~\ref{sec:conclusionsAndFutureWork}).
\end{itemize}

\section{The LQR Model and Preliminaries}\label{sec:lqrModelPreliminaries}

The state of the system under study at time $t\in\mathsf{N} = \{0,1,\cdots\}$, is denoted by $\x_t$ and evolves in $\reals^d$ according to the following law
\begin{equation}
\x_{t+1} = A\x_{t}+\b u_t+ \mathbf{c}_t + \mathbf{n}_t,t\in\mathsf{N}. \tag{SYS}
\label{eqn:LinearSystemWithConstForcingFunction}
\end{equation}

We now proceed to explain the various elements of the above evolution equation.  Here, $\u_t\in\reals$ is the control input deployed in time slot $t,$ and $\mathbf{n}_t\in\reals^d$ models random state disturbance. The system gain and control coefficient matrices are denoted by $A\in\reals^{d\times d}$ and $\b\in\reals^{d\times 1},$ respectively and $\mathbf{c}$ is a constant activation. The state disturbance process $\left\lbrace\mathbf{n}_t\right\rbrace_{t\in\mathsf{N}}$ comprises IID random vectors drawn from some distribution $F_n$ over $\reals^d$ with $\E\mathbf{n}_t=\boldsymbol{0}$ and $\Psi_n := \E\mathbf{n}_t\mathbf{n}^T_t$. Single stage cost is defined by 
\begin{eqnarray}
g(\x_t,u_t) &:=& \underbrace{\x^T_tQ\x_t}_{\text{state penalizing}}+\underbrace{{\color{blue}r}~u^2_t}_{\text{control penalizing}},~t\in\mathsf{N},
\label{eqn:quadraticCostWithControlPenalty}
\end{eqnarray}
where the matrix $Q$ is assumed to be symmetric positive semidefinite (SPSD). 
Clearly, the first term in Eqn.~\ref{eqn:quadraticCostWithControlPenalty} (i.e., $\x^T_t Q\x_t$) penalizes the state while the latter penalizes the control input $u_t$. The control penalizing parameter (or control penalty coefficient) $r>0$ in \eqref{eqn:quadraticCostWithControlPenalty} will be of particular interest to us in subsequent sections. Given a discount factor $\gamma\in(0,1),$ the control input is to be designed to minimize a $\gamma$-discounted cost defined by (we suppress the dependence of $J_r(\cdot)$ on $\gamma$ for ease of exposition)
\begin{eqnarray}
    J_r(\x_0) &:=& \mathbb{E}\sum_{t=0}^\infty \gamma^t g(\x_t,u_t)\nonumber\\
              &=& \mathbb{E}\sum_{t=0}^\infty \gamma^t \left(\x^T_tQ\x_t + r~u^2_t\right),
              \label{eqn:gammaDiscountedCostCENTRALIZEDLinearQuadraticSystemWithControlPenalty}
\end{eqnarray}
where the expectation is over both the state disturbance and the (possibly randomized) controls $\{u_t\}_{t=0}^\infty.$ In \eqref{eqn:gammaDiscountedCostCENTRALIZEDLinearQuadraticSystemWithControlPenalty}, we have made explicit the fact that the cost functional $J_r(\cdot)$ depends on the control penalizing parameter $r$ in anticipation of results to follow that will explore the impact of control penalization on various performance metrics. Note that the tuple $(A,\b,\mathbf{c},F_n,Q,r,\gamma)$ completely characterizes this LQR system. For simplicity of exposition, we will assume $\mathbf{c}_t=\mathbf{0}$ in the sequel and parameterize the system by $\color{blue}(A,\b,F_n,Q,r,\gamma)$ without loss of generality. We denote by $\mathcal{P}(\mathbb{R})$ the set of all probability distributions on $\mathbb{R}.$ 

For every $t\in\mathsf{N},$ let $\mathcal{F}_t$ denote the sigma algebra generated by $\left\lbrace \x_s,u_s,~s< t\right\rbrace \cup \{\x_t\}$. Throughout the article we will only work with nonanticipatory controls, i.e., those that, at time $t,$ use only the information available up to $t$ and are of the form $u_t: \mathcal{F}_t\rightarrow\mathcal{P}(\mathbb{R})$. 
Restricting our attention to Markov controls, $u_t = \mu_t(\x_t),~t\in\mathsf{N}$, where $\mu_t:\reals^d\rightarrow\mathcal{P}(\mathbb{R})$ we define a  {\color{blue}(Markov) policy} $\pi$ to be a time indexed sequence of such maps $\pi = \left[\mu_0,\mu_1,\cdots\right]$. If the  constituent maps are independent of time, the control policy is said to be \emph{\color{blue}stationary}, in which case $\pi = \left[\mu,\mu,\cdots\right]$. We denote by $\Pi$ the set of all (nonanticipatory) policies for this system, by $\mathbb{F}\subset\Pi$ the set of all stationary policies and $\mathbb{D}\subset\mathbb{F}$, the set of all stationary \emph{\color{blue}deterministic} policies. 

The probability measure on the space of sample paths induced by policy $\pi$ beginning in state $\x_0$ is denoted by $\mathbb{P}^{\pi}_{\x_0}$ and the associated expectation operator by $\mathbb{E}^{\pi}_{\x_0}$. The optimal cost functional is defined by 
\begin{eqnarray}
    J_r^*(\x_0) := \inf_{\pi\in\Pi} \mathbb{E}^{\pi}_{\x_0}\sum_{t=0}^\infty \gamma^t \left(\x^T_tQ\x_t + r~u^2_t\right).
              \label{eqn:OptimalGammaDiscountedCostCENTRALIZEDLinearQuadraticSystemWithControlPenalty}
\end{eqnarray}
Using Propositions~3.1.3 and 3.1.5 in \cite{bertsekas11dynamic-programming-optimal-control-volume2} we see that the optimal policy $\pi^* = \left[\mu^*,\mu^*,\cdots\right]$ for this system is stationary and given by 
\begin{equation}
    u^*_t = \mu^*(\x) = -\gamma\frac{1}{\gamma\b^t K \b+r} \b^t K A \x,
\label{eqn:optimalCENTRALIZEDControlLQGSystem}
\end{equation}
where $K$ is an SPSD matrix that satisfies the recursion
\begin{equation}\label{eqn:DTRiccatiEquation}
    K  = Q + A^T \left[\gamma K - \frac{1}{\gamma\b^T K \b+r}\gamma^2 K \b\b^T K\right] A.
\end{equation}
A unique solution to \eqref{eqn:DTRiccatiEquation} exists (and is, in fact, positive \emph{definite}) provided the pair $(A,b)$ is controllable and the pair $(A,C)$ is observable \cite[Prop.~4.1]{bertsekas95dynamic-programming-optimal-control-volume1}, where $C = \sqrt{\Lambda}M$ and $\Lambda$ and $M$ are diagonal and orthogonal matrices respectively such that $Q = M^T\Lambda M$. Therefore, for the rest of the paper we have
\begin{asmpn}\label{assumption:controllabilityAndObservability}
The system $(A,\b)$ is controllable and the pair $(A,C)$ is observable\footnote{In fact, observability is easily guaranteed by choosing $Q$ to be positive definite, which can be ensured by minimally perturbing the chosen $Q$ with a scaled identity matrix.}.
\end{asmpn}
Under Assumption~\ref{assumption:controllabilityAndObservability}, the optimal cost is given by 
\begin{equation}\label{eqn:optimalCostSystemWithControlPenalty}
    J^*_r(\x_0) = \x_0^TK\x_0 + \frac{\gamma}{1-\gamma}\mathbb{E}\mathbf{n}^TK\mathbf{n},~\forall\x_0\in\mathbf{R}^3,
\end{equation}
where $\mathbf{n}\sim F_n$, the state disturbance distribution. 

Given any pricing policy $\pi:=[u_0,u_{1},\cdots]$ for the above system, define the {\color{blue}Volatility} of a policy by
\begin{eqnarray}\label{eqn:defnOfPolicyVolatility}
\mathcal{V}_{\x_0}(\pi) := \mathbb{E}^{\pi}_{\x_0}\sum_{t=0}^{\infty}\gamma^t u^2_t.
\end{eqnarray}
 Also define {\color{blue}Regulator Efficiency} or \emph{System Efficiency} under $\pi$ by
\begin{eqnarray}\label{eqn:defnOfPolicyEfficiency}
\mathcal{E}_{\x_0}(\pi) := -\mathbb{E}^{\pi}_{\x_0}\sum_{t=0}^{\infty}\gamma^t \x_t^TQ\x_t.
\end{eqnarray}
We will call a control policy $\pi\in\Pi$ \emph{\color{blue}admissible} if it the cost and volatility functionals are well defined under $\pi$ and denote by $\mathcal{A}\subset\Pi$ the subset of all admissible policies. Finally, we define a \emph{\color{blue}Bellman Operator} as follows. Let $\mathcal{S} = \reals^d\times(0,\infty)$ and let $\mathcal{J} = \left\lbrace v\mid v:\mathcal{S}\rightarrow \mathbb{R}\right\rbrace$ be the space\footnote{We use \enquote{space} instead of \enquote{set} because we will convert $\mathcal{J}$ to a metric space shortly.} of all functionals on $\mathcal{S}$. The Bellman Operator $T:\mathcal{J}\rightarrow\mathcal{J}$ is given by
\begin{equation}\label{eqn:BellmanOperator}
    (Tv)(\x,r) := \x^TQ\x + \inf_{u\in\mathbb{R}} r~u^2 + \gamma \mathbb{E}_{\mathbf{n}}v(A\x+\b u+\mathbf{n},r).
\end{equation}
Having constructed the requisite theoretical scaffolding we will now move on to deriving a formal relationship between control fluctuations and system efficiency. Please note that we have provided a {\color{blue}glossary of notation} for the convenience of the reader in Sec.~\ref{sec:appendixGlossaryNotation} in the Appendix.

\section{Regulator Efficiency vs Control Volatility}\label{sec:regulatorEffVersusControlVolatilityinLQR}
We begin this section with Thm.~\ref{thm:CostWithControlPenaltyConcaveInR} wherein we show that the optimal cost function, $J^*_r$, satisfies two important properties. 
Aided by this result, we then show that there exists a fundamental tradeoff between the volatility and efficiency functionals defined in Eqns.~\eqref{eqn:defnOfPolicyVolatility} and \eqref{eqn:defnOfPolicyEfficiency} under the control policy $\pi^*$ that achieves the optimal cost \eqref{eqn:OptimalGammaDiscountedCostCENTRALIZEDLinearQuadraticSystemWithControlPenalty}. Next, invoking Thm.~\ref{thm:CostWithControlPenaltyConcaveInR}, we show that there exists a fundamental relation between the system efficiency that a control policy can achieve and its volatility. We also characterize the space of all (volatility,efficiency) pairs that can be achieved by any admissible control mechanism in a given LQR $(A,\b,F_n,Q,r,\gamma)$. 

This latter result is of practical importance since it can be used to inform the controller about the consequences of policy decisions on system efficiency. On the flip side, it is also useful to know in advance, how much price volatility one can expect while attempting to achieve a desired efficiency level. We now discuss two important properties of $J^*_r.$
\begin{tcolorbox}
\begin{thm}\label{thm:CostWithControlPenaltyConcaveInR}
$J_r^*(\x)$ is concave and nondecreasing in $r.$ Specifically,
\begin{equation}
\begin{aligned}
\frac{dJ^*_{r}(\x)}{dr} &> 0,\text{ and }\\
\frac{d^2J^*_{r}(\x)}{dr^2} &\leq 0,~\forall \x\in\mathbb{R}^3.\label{eqn:CostWithControlPenalty1stAnd2ndDerivatives}
\end{aligned}
\end{equation}
\end{thm}

\end{tcolorbox}
\begin{pf}
The proof proceeds in several steps. We first define a norm $\|\|:\mathcal{J}\rightarrow\mathbb{R}_+$ by 
\begin{equation}\label{eqn:supNormOnFunctionSpaces}
    \|v\| := \sup_{\mathbf{y}\in\mathcal{S}}\frac{|v(\mathbf{y})|}{\left(\|\mathbf{y}\|_2\vee1\right)^2},
\end{equation}
and let $\mathcal{V}:=\{v\in\mathcal{J}:\|v\|<\infty\}$. 
Clearly, $J^*_r\in\mathcal{V}$. This norm induces a metric $\rho:\mathcal{V}\times\mathcal{V}\rightarrow\mathbb{R}_+$ defined by $\rho(v,w) = \|v-w\|$. Before proceeding further we first note that 
\begin{lem}\label{lem:VRhoIsACompleteMetricSpace}
$\langle V, \rho \rangle$ is a complete metric space, and $v\in\mathcal{V}\Rightarrow T v\in\mathcal{V}$, i.e., the Bellman operator preserves finiteness of $\parallel\parallel.$
\end{lem}
\begin{pf}
See Sec.~\ref{app:ProofOflem:VRhoIsACompleteMetricSpace} in the Appendix.
\end{pf}

Next, we characterize the subspace of all $v\in\mathcal{V}$ that are nondecreasing and concave.
\begin{lem}\label{lem:SubspaceHClosedUnderT}
Let $\mathcal{H} := \{v\in\mathcal{V}: v\text{ is concave and nondecreasing}\}$. Then, 
\begin{enumerate}
    \item $\mathcal{H}$ is a closed subset of $\mathcal{V}$, where closure is in the topology induced by $\rho$, and 
    \item $v\in\mathcal{H}\Rightarrow Tv\in\mathcal{H}.$
\end{enumerate}
\end{lem}
\begin{pf}
See Sec.~\ref{app:ProofOflem:SubspaceHClosedUnderT} in the Appendix.
\end{pf}

We now note that (a) the function $v_0(\mathbf{y}) := 0,~\forall\mathbf{y}\in\mathcal{S}$ is in $\mathcal{H}$, (b) that $J^*$ is a fixed point of the operator $T$ and observe the following
\begin{lem}\label{lem:zeroFunctionConvergesToJStar}
$\lim_{k\rightarrow\infty}T^kv_0 = J^*$, where convergence is in the topology induced by $\rho$.
\end{lem}
\begin{pf}
See Sec.~\ref{app:ProofOflem:zeroFunctionConvergesToJStar} in the Appendix.
\end{pf}

The above lemma, together with the fact that $\mathcal{H}$ is closed, means that $J^*\in\mathcal{H}.$ Furthermore, it is evident that $r\mapsto J^*_r(\x)\in\mathcal{C}^2\left((0,\infty)\right),$ whence we have \eqref{eqn:CostWithControlPenalty1stAnd2ndDerivatives}. This concludes the proof. 
%
\end{pf}

The simulation result in Fig.~\ref{fig:optimalAndStatePenalizingCostsConcaveInR} shows the concavity of $J^*_r$ for the market specified in Sec.~\ref{app:detailsOfeExperiments}. Now,  Thm.~\ref{thm:CostWithControlPenaltyConcaveInR} has several consequences. To begin with, we show in Sec.~\ref{sec:OptimalEfficiencyVolatilityTradeoff} that it implies a fundamental tradeoff between Efficiency and Volatility for policy $\pi^*$ defined in \eqref{eqn:optimalCENTRALIZEDControlLQGSystem}. Specifically, we will now show that the efficiency functional with $\pi^*,$ i.e., $\mathcal{E}_{\x_0}(\pi^*)$
decreases with decreasing volatility, which means that improving optimal system efficiency always comes at the cost of increased control volatility. However, in Sec.~\ref{sec:CapacityRegionNoFreeLunchTheorem} we go even further and prove a stronger claim that this volatility-efficiency tradeoff must be satisfied by every admissible control policy. 

\subsection{The Volatility-Efficiency tradeoff}\label{sec:OptimalEfficiencyVolatilityTradeoff}
Recall that the first term in the single stage cost was called the \emph{state penalizing} portion (see Eqn.~\eqref{eqn:quadraticCostWithControlPenalty}) and consider the impact of employing the policy given in \eqref{eqn:optimalCENTRALIZEDControlLQGSystem} on this cost alone, i.e., on 
\begin{eqnarray}
  J_{sp,r}(\x_0) := \mathbb{E}\sum_{t=0}^\infty \gamma^t\x^T_tQ\x_t = -\mathcal{E}_{\x_0}(\pi^*),
\label{eqn:gammaDiscountedCostCENTRALIZEDLinearQuadraticSystemNoControlPenalty}
\end{eqnarray}
where the subscript $sp$ stands for \enquote{state penalizing.} We show, in Prop.~\ref{prop:discountedCostApplyingUStarWithoutControlPenalty} in the Appendix, that this cost is given by
\begin{eqnarray}
  J_{sp,r}(\x_0) = \x^T_0S\x_0 + \frac{\gamma}{1-\gamma}\mathbb{E}\mathbf{n}^TS\mathbf{n},
\label{eqn:gammaDiscountedNoControlPenaltyInTermsOfS}
\end{eqnarray}
where $S$ is an SPSD matrix that satisfies
\begin{eqnarray}
S = Q + \gamma \left(A^T \left[ I - \gamma\frac{1}{\gamma\b^T K \b+r} K\b\b^T \right] S \left[ I - \gamma\frac{1}{\gamma\b^T K \b+r} \b\b^T K\right] A\right).
\label{eqn:riccatiEqnForStatePenalizingCost}
\end{eqnarray}
\begin{prop}\label{prop:existenceOfSolutionToRiccatiRecursionWithoutControlCost}
Under Assumption~\ref{assumption:controllabilityAndObservability}, a unique symmetric positive definite matrix satisfying \eqref{eqn:riccatiEqnForStatePenalizingCost} exists. Moreover, the sequence of matrices generated by 
\begin{eqnarray}
    S_{t+1} &=& Q + \gamma \left(A^T \left[ I - \gamma\frac{1}{\gamma\b^T K \b+r} K\b\b^T \right]S_t \left[ I - \gamma\frac{1}{\gamma\b^T K \b+r} \b\b^T K\right] A\right).
  \label{eqn:recursion4riccatiEqnForStPenCost}
\end{eqnarray}
with $S_0$ set to any SPSD matrix, 
satisfies
\begin{eqnarray}
  \lim_{t\rightarrow\infty} \parallel S_t-S\parallel_F=0,
\end{eqnarray}
where $S$ is the solution of \eqref{eqn:riccatiEqnForStatePenalizingCost} and $\parallel M\parallel_F$ is the Frobenius norm of matrix $M.$
\end{prop}
\begin{pf}
We first show in Prop.~\ref{prop:discountedCostApplyingUStarWithoutControlPenalty} in the Appendix, that $J_{sp,r}$ satisfies a Bellman-like fixed point equation which then gives us \eqref{eqn:gammaDiscountedNoControlPenaltyInTermsOfS} . Thereafter, we modify a general technique used in the analysis of Riccatti equations to first prove convergence in the Frobenius norm of \eqref{eqn:recursion4riccatiEqnForStPenCost} when initialized with $S_0 = [0]_{3\times3}$ and extend the result to $S_0$ being any arbitrary SPSD matrix. Finally, we show that the limit point has to be unique.
See Sec.~\ref{app:ProofOfPropExistSolRiccatiWithoutCtrlCost} in the Appendix for details.
\end{pf}

Returning to \eqref{eqn:gammaDiscountedNoControlPenaltyInTermsOfS}, we now show that similar to our original cost functional $J^*_r(),$ the state penalizing functional is also concave nondecreasing in $r,$ that is
\begin{tcolorbox}

\begin{thm}[V-E tradeoff]\label{thm:CostWithOutControlPenaltyConcaveInR}
$J_{sp,r}(\x)$ is concave nondecreasing in $r.$ Specifically,
\begin{equation}
\begin{aligned}
\frac{dJ_{sp,r}(\x)}{dr} &> 0,\text{ and }\\
\frac{d^2J_{sp,r}(\x)}{dr^2} &\leq 0,~\forall \x\in\mathbb{R}^3.\label{eqn:CostWithoutControlPenalty1stAnd2ndDerivatives}
\end{aligned}
\end{equation}
\end{thm}

\end{tcolorbox}

\begin{pf}
The proof uses the ideas in the proof of Thm.~\ref{thm:CostWithControlPenaltyConcaveInR}, along with the fact that beginning with any SPSD matrix, the iterates in \eqref{eqn:recursion4riccatiEqnForStPenCost} converge in the Frobenius norm. Details can be found in Sec.~\ref{app:ProofOfCostWithoutControlPenaltyConcave} in the Appendix.
\end{pf}

Since state penalizing cost is defined to be the negative of the efficiency functional in Eqn.~\eqref{eqn:gammaDiscountedCostCENTRALIZEDLinearQuadraticSystemNoControlPenalty}, Thm.~\ref{thm:CostWithOutControlPenaltyConcaveInR} shows that efficiency increases with increasing volatility, establishing the volatility-efficiency tradeoff. We thus see that operating an LQR at high efficiency automatically entails highly volatile control inputs. The simulation results presented in Sec.\ref{sec:simulationResults} illustrate this phenomenon very clearly.

\section{The Capacity Region for Linear Quadratic Regulators and A \enquote{No Free Lunch} theorem}\label{sec:CapacityRegionNoFreeLunchTheorem}
The discussion hitherto centered around \emph{optimal} control inputs, which naturally indicates the possibility of the existence of such a relationship for other control policies. In this section, we answer that question in the affirmative. Suppose we impose upon the stochastic control problem described in Sec.~\ref{sec:lqrModelPreliminaries}, the constraint that the volatility of the control law must not exceed a certain threshold $\alpha>0$, i.e., that 
\begin{equation}
    \mathcal{V}_{\x_0}(\pi) = \mathbb{E}^{\pi}_{\x_0}\sum_{t=0}^{\infty}\gamma^t u^2_t \leq \alpha,~\forall\x_0\in\reals^d.
\end{equation}
For a given level $\alpha>0,$ we define $\mathcal{A}_{\alpha}\subset\mathcal{A}$ to be the set of all policies whose volatility is less than $\alpha$ and call these policies \emph{\color{blue}$\alpha$-admissible.} 
The policy that attains maximum efficiency, if it exists, must satisfy
\begin{eqnarray}
  \argmax_{\pi\in\Pi} && -\mathbb{E}^{\pi}_{\x_0}\sum_{t=0}^\infty \gamma^t\x^T_tQ\x_t \nonumber\\
  \text{subject to }  && \x_{t+1} = A\x_{t}+\b u_t+\mathbf{n}_t,t\in\mathsf{N}\text{, and} \label{eqn:marketEfcyConstrainedMDPFormulation}\\
                      && \mathbb{E}^{\pi}_{\x_0}\sum_{t=0}^\infty \gamma^t u_t^2 \leq \alpha.\nonumber
\end{eqnarray}
Clearly, all feasible solutions to the above problem are $\alpha$-admissible and hence, the optimizer always lies within $\mathcal{A}_\alpha.$
This problem can be recast in the form of a \emph{constrained} Markov decision process  \cite{altman99constrained-markov-decision-processes,lopez-lerma03lagrange-approach-constrained-mdps-survey-extension}, and solved with the aid of  the Lagrangian $\mathcal{L}:\Pi\times(0,\infty)\rightarrow\mathbb{R}$, defined by
\begin{eqnarray}
    \begin{aligned}
        \mathcal{L}(\pi,\lambda) &:=& \mathbb{E}^{\pi}_{\x_0}\sum_{t=0}^\infty \gamma^t \x^T_tQ\x_t+ \lambda \left( \mathbb{E}^{\pi}_{\x_0}\sum_{t=0}^{\infty}\gamma^t u^2_t - \alpha \right)\\
                              &=& \mathbb{E}^{\pi}_{\x_0}\sum_{t=0}^\infty \gamma^t \left(\x^T_tQ\x_t + \lambda~u^2_t\right) - \lambda\alpha, 
                                \label{eqn:LagrangianCMDP4VETradeoff}
    \end{aligned}
\end{eqnarray}
where we deliberately suppress the dependence of $\mathcal{L}$ on $\x_0$ for ease of exposition. Note that in \eqref{eqn:LagrangianCMDP4VETradeoff} the Lagrangian represents a cost minimization problem rather than an efficiency maximization problem. So our aim now will be to \emph{minimize} this Lagrangian.
Invoking Theorems~4.2 and 4.4 in \cite{lopez-lerma03lagrange-approach-constrained-mdps-survey-extension}, we see that the optimal cost $\mathcal{L}^*$ for this constrained problem is given by 
\begin{eqnarray}
\hspace{-0.5cm}
  \mathcal{L}^* &=& \inf_{\pi\in\Pi} \sup_{\lambda\geq0} \mathcal{L}(\pi,\lambda) = \sup_{\lambda\geq0} \inf_{\pi\in\Pi} \mathcal{L}(\pi,\lambda)\nonumber\\
    &=& \sup_{\lambda\geq0} \left(\inf_{\pi\in\Pi} \mathbb{E}^{\pi}_{\x_0}\sum_{t=0}^\infty \gamma^t \left(\x^T_tQ\x_t + \lambda~u^2_t\right)\right) - \lambda\alpha. 
    \label{eqn:optimalCostOfConstrainedMDP}
\end{eqnarray}
We solve for $\mathcal{L}^*$ using this last expression. Notice that keeping $\lambda\geq0$ fixed and using \eqref{eqn:optimalCostSystemWithControlPenalty}, $\inf_{\pi\in\Pi} \mathcal{L}(\pi,\lambda) = \x_0K_\lambda\x_0 + \frac{\gamma}{1-\gamma}\mathbb{E}\mathbf{n}^TK_\lambda\mathbf{n},~\forall\x_0\in\reals^d,$ where 
\begin{eqnarray}
  K_\lambda = Q + A^T \left[\gamma K - \frac{1}{\gamma\b^t K \b+\lambda}\gamma^2 K \b\b^T K\right] A.
\end{eqnarray}
This gives us $\mathcal{L}^* = \sup_{\lambda\geq0}~\x_0K_\lambda\x_0 + \frac{\gamma}{1-\gamma}\mathbb{E}\mathbf{n}^TK_\lambda\mathbf{n}-\lambda \alpha,$ where $\mathbf{n}\sim F_n,$ the state disturbance distribution. At the optimal $\lambda,$ we have $\x_0\frac{dK_\lambda}{d\lambda}\x_0 + \frac{\gamma}{1-\gamma}\mathbb{E}\mathbf{n}^T\frac{dK_\lambda}{d\lambda}\mathbf{n} = \alpha,$ but since $\frac{d^2K_\lambda}{d\lambda^2} \leq 0$ (non positive definite), as $\alpha$ decreases, i.e., as we demand lesser control volatility, the optimal value of $\lambda$ increases which, as Thm.~\ref{thm:CostWithControlPenaltyConcaveInR} shows, means that $\x_0K_\lambda\x_0+ \frac{\gamma}{1-\gamma}\mathbb{E}\mathbf{n}^TK_\lambda\mathbf{n}$ increases which results in decreased system efficiency.

However, we can go even further and make a stronger claim. We have already shown that the optimal cost of the constrained MDP, $\mathcal{L}^*,$ is nonincreasing in the price volatility $\alpha.$ 
Notice that the Lagrangian dual, i.e.,
\begin{eqnarray}
\mathcal{L}^* &=& \sup_{\lambda\geq0}~\x_0K_\lambda\x_0 + \frac{\gamma}{1-\gamma}\mathbb{E}\mathbf{n}^TK_\lambda\mathbf{n}-\lambda \alpha,
        \label{eqn:optimalLagrangianSupOverLambdaConstrainedMDP}
\end{eqnarray}
is convex in $\alpha.$ 
Therefore, optimal market efficiency $\mathcal{E}^*_{\x_0} = -\mathcal{L}^*$ is \emph{concave} in the  volatility level $\alpha$ (Fig.~\ref{fig:marketCapacityRegionSimulated} helps validate this
conclusion experimentally for regulator whose specifications are provided in Sec.~\ref{app:detailsOfeExperiments} in the Appendix). Since $\mathcal{E}^*_{\x_0}$ is the maximum achievable efficiency for any given $\alpha,$ any other $\alpha$-admissible policy $\pi\in\mathcal{A}_\alpha$ will necessarily only achieve lower (or at the very least not higher) efficiency. The foregoing arguments appear to suggest that higher efficiency efficiency levels cannot be attained by any policy without sacrificing price volatility. We now show that this intuition is correct.
Consider the tuple $\left(\mathcal{V}_{\x_0}(\pi),\mathcal{E}_{\x_0}(\pi)\right)$ for every $\pi\in\mathcal{A}_\alpha$ and let 
\begin{eqnarray}
    \mathcal{C}_\alpha &:=& \left\lbrace\left(\mathcal{V}_{\x_0}(\pi),\mathcal{E}_{\x_0}(\pi)\right):\pi\in\mathcal{A}_\alpha\right\rbrace\nonumber.
\end{eqnarray}
We then have the following \enquote{no free lunch} theorem.
\begin{tcolorbox}
\begin{thm}[LQR Capacity Region]\label{thm:CapacityRegionOfMarket}
For every LQR $(A,\b,F_n,Q,r,\gamma),$ there exists a set $\mathcal{C}\subset\mathbb{R}^2$ 
\begin{equation}
    \mathcal{C} = \bigcup_{\alpha>0} \mathcal{C}_\alpha,
\end{equation}
such that 
\begin{itemize}
    \item[a.] $\mathcal{C}$ is closed and convex,
    \item[b.] the Pareto boundary of $\mathcal{C}$ is given by the optimal efficiency $\mathcal{E}^*_{\x_0},$ and
    \item[c.] for every $\pi\in\mathcal{A},$ the tuple $\left(\mathcal{V}_{\x_0}(\pi),\mathcal{E}_{\x_0}(\pi)\right)\in\mathcal{C}$.
\end{itemize}
\end{thm}
\end{tcolorbox}
\begin{pf}
The proof essentially involves showing that the set of all achievable (volatility, efficiency) tuples is contained in $\mathcal{C}$ and later, that every point in $\mathcal{C}$ is achievable. Details are presented in Sec.~\ref{app:ProofOfThmCapacityRegionOfMarket} in the Appendix.
\end{pf}

Fig.~\ref{fig:marketCapacityRegion} illustrates the set $\mathcal{C}$ and the Pareto boundary for a given regulator $(A,\b,F_n,Q,r,\gamma)$. Since this region completely determines all achievable efficiencies under every acceptable level of price volatility, we call $\mathcal{C}$ the \textbf{\color{blue}Capacity Region} of the market $(A,\b,F_n,Q,r,\gamma)$, similar to the Shannon capacity region for the physical layer \cite{cover-thomas12elements-information-theory} and the Tassiulas-Ephremides capacity region for the MAC layer \cite{tassiulas-ephremides90stability-constrained-queueing-multihop} of a communication system. An example of such a region is shown through simulations in Fig.~\ref{fig:marketCapacityRegionSimulated} in Sec.~\ref{sec:simulationResults} for a particular regulator. This region is especially useful for estimating an upper bound on the efficiency achievable both by different regulators and by various control mechanisms for a given system.

In what follows, we discuss the consequences of the results in the foregoing sections on deregulated electricity markets, which will help concretely illustrate their impact in real-world scenarios. We begin Sec.~\ref{sec:marketsWithPriceAnticipatingParticipants} and Sec.~\ref{sec:applicationPriceTakingMarketAnalysis} by deriving a model for such a market.This will help translate Theorems~\ref{thm:CostWithOutControlPenaltyConcaveInR} and \ref{thm:CapacityRegionOfMarket} to the deregulated market setting. Thereafter, the same  analysis will also help explain certain volatility phenomena observed in regions with high penetration of renewable energy sources. 
\begin{figure}[htb]
\centering
\tikzset{every picture/.style={line width=0.75pt}} 

\begin{tikzpicture}[x=0.5pt,y=0.5pt,yscale=-1,xscale=1]
\filldraw[very thick, fill=blue!60!white!40, draw=black] (319.2,330.8) .. controls (350.4,58.3) and (569.38,102.6) .. (620.38,97.6) ;

\filldraw[blue!60!white!40] (319.2,330.8) -- (620.38,97.6) -- (620.38,330.8) -- cycle; 

\draw[thick,black]  (274.4,330.8) -- (652.4,330.8)(319.2,80.6) -- (319.2,358.6) (645.4,325.8) -- (652.4,330.8) -- (645.4,335.8) (314.2,87.6) -- (319.2,80.6) -- (324.2,87.6)  ;

\draw[thick,black,dashed]   (425,131) -- (428.35,329.6) ;
\draw [shift={(428.38,331.6)}, rotate = 269.03] [color={rgb, 255:red, 0; green, 0; blue, 0 }  ][line width=0.75]    (10.93,-3.29) .. controls (6.95,-1.4) and (3.31,-0.3) .. (0,0) .. controls (3.31,0.3) and (6.95,1.4) .. (10.93,3.29)   ;

\draw[thick,black,dashed]   (425,131) -- (355.38,131.32) -- (319.38,132.53) ;
\draw [shift={(317.38,132.6)}, rotate = 358.06] [color={rgb, 255:red, 0; green, 0; blue, 0 }  ][line width=0.75]    (10.93,-3.29) .. controls (6.95,-1.4) and (3.31,-0.3) .. (0,0) .. controls (3.31,0.3) and (6.95,1.4) .. (10.93,3.29)   ;

\draw[thick,black]    (553.38,99.6) .. controls (582.24,137.41) and (589.69,81.76) .. (627.81,137.75) ; 
\draw [shift={(628.38,138.6)}, rotate = 236.07] [color={rgb, 255:red, 0; green, 0; blue, 0 }  ][line width=0.75]    (10.93,-3.29) .. controls (6.95,-1.4) and (3.31,-0.3) .. (0,0) .. controls (3.31,0.3) and (6.95,1.4) .. (10.93,3.29)   ; 


\draw (266,205) node  [rotate=-269.64] [align=left] {Efficiency $\displaystyle (\mathcal{E}_{\mathbf{x}}( \pi ))$};
\draw (545,354) node   [align=left] {Volatility $\displaystyle (\mathcal{V}_{\mathbf{x}}( \pi ))$};
\draw (426,341) node   [align=left] {$\displaystyle \alpha$};
\draw (306,137) node  [rotate=-270.66] [align=left] {$\displaystyle \mathcal{E}^{*}_{\x}$ at $\displaystyle \alpha$};
\draw (651,149) node   [align=left] {\bf Pareto boundary};
\draw (523,226) node   [align=left] {\bf Capacity Region $\displaystyle \mathcal{C}$};

\end{tikzpicture}
\caption{The Capacity Region of the regulator $(A,\b,F_n,Q,r,\gamma)$. The Pareto boundary is the maximum efficiency, $\mathcal{L}^*$, of the constrained MDP \eqref{eqn:optimalCostOfConstrainedMDP}. No $(\mathcal{V},\mathcal{E})$ pair outside $\mathcal{C}$ is achievable in this system.}
\label{fig:marketCapacityRegion}
\end{figure}
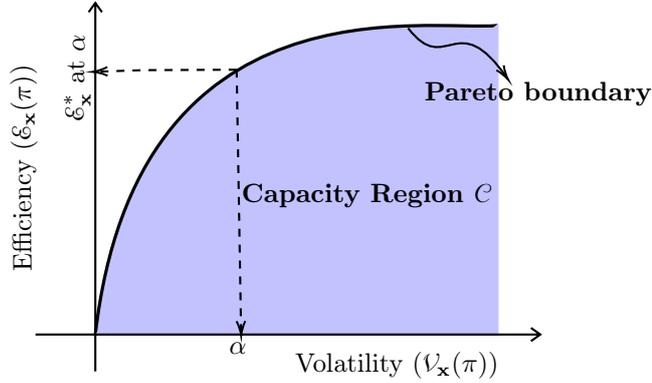


\section{Applications to the Analysis of Deregulated Electricity Markets, Part I: Price Anticipating Participants}\label{sec:marketsWithPriceAnticipatingParticipants}
As mentioned in Sec.~\ref{sec:Introduction}, deregulated electricity markets in Europe and the United States are managed by two types of \emph{profit-neutral} entities called Independent System Operators (ISOs) and Regional Transmission Organizations (RTOs). RTOs and ISOs manage the power grid via two sub-markets: Day Ahead Markets and Real Time Markets that operate on time scales that range from 24 hours to 5 minutes. Electricity markets, therefore, are naturally amenable to being modeled as time-slotted systems. It is to be noted that here, the ISO or the RTO plays the role of a coordinating agent and, being profit-neutral, is only allowed to control the system through the price of electricity that it is tasked with setting. This technique of using of market price as a coordinating mechanism is grounded in well-established microeconomic theory and began with the seminal work of Walras \cite{walras13elements-pure-economics}. There have since been multiple attempts to apply this theory to analyze electricity markets \cite{hogan92contract-networks-electric-transmission,caramanis-etal82optimal-spot-pricing-practice-theory,bohn-caramanis84optimal-pricing-space-time}.

\subsection{Modeling the Power Market}\label{sec:priceAnticipatingMarketMODEL}
The electric power market has $N_c,N_p$ consumers and producers respectively, who are generically termed \emph{prosumers.} In each time slot, every market participant submits bids in the form of \emph{Price-Quantity graphs} (PQ graphs), also termed \emph{Inverse Demand/Cost Functions} \cite{mas-collel-etal95microeconomic-theory} to the RTO or the ISO as the case may be. The market operator then determines and publishes: (1) the clearing price of electricity (also called the \emph{equilibrium} price), and (2) how much each prosumer is allowed to produce/purchase as the case may be, in that time slot. This market is, therefore, a perpetually repeating sequence of auctions. However, what complicates the analysis is the fact that future demand and supply are coupled with the price process which is itself, a function of the submitted PQ graphs. 

\begin{figure}[htb]
\centering
\input{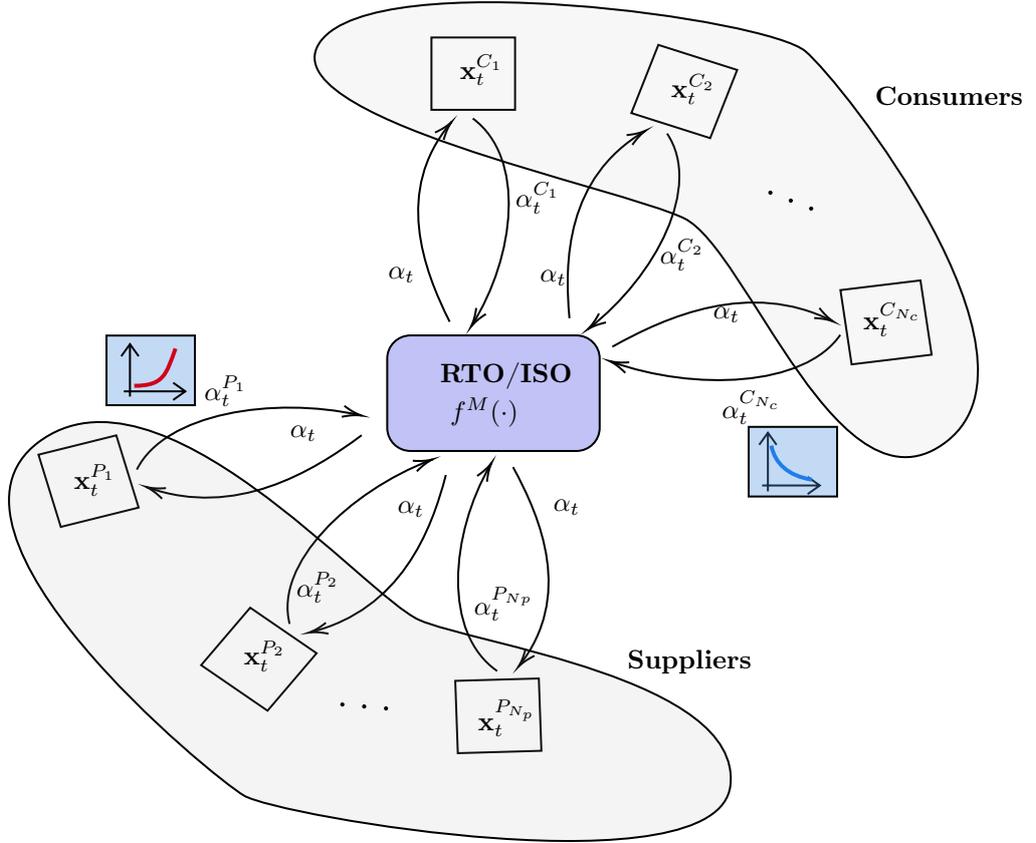}
\caption{The operation of the electric power market with strategic prosumers. At time $t\geq0,$ $\xc$ is the state of Consumer~$i,~i\in[N_c]$ and $\xp$ is the state of Supplier~$j,~j\in[N_p]$. Market participants submit bids in the form of PQ graphs (depicted in the blue-shaded boxes) parameterized by $(\alpha^{c_i}_t)_{i\in[N_c]},(\alpha^{p_j}_t)_{j\in[N_p]},$ and the RTO/ISO decides the clearing price $\alpha_t.$ }
\label{fig:price-anticipating-market}
\end{figure}

As Fig.~\ref{fig:price-anticipating-market} shows, the state of every consumer $c_i,~i\in[N_c]$ at time $t\in\mathsf{N}=\{0,1,\cdots\}$ is modeled by the 3-tuple $\xc_t:=\left[d^i_t,s^{c_i}_t,\alpha^{c_i}_t\right]\in\mathbb{R}^3$, where $d^i_t$ is the quantity demanded by consumer $c_i$, $s^{c_i}_t$ is the power allocated to it and $\alpha^{c_i}_t$ parameterizes its pre-announced PQ graph function $q^{c_i}_{\alpha^{c_i}_t}(\cdot)$. We assume here that the \emph{form} of each of the $N_c$ price-quantity graphs, each of which is a mapping $q^{c_i}_{\alpha}:\mathbb{R}_+\rightarrow\mathbb{R}_+$ is common knowledge, that only the specific parameter $\alpha$ is unknown and this parameter is strategically set by each consumer in every time slot. We will make a similar assumption about the suppliers' PQ graphs. Similarly, the state of every producer $p_i,~j\in[N_p]$ is modeled by $\xp_t:=\left[s^j_t,\alpha^{p_j}_t\right]\in\mathbb{R}^2$, where $s^j_t$ is the quantity supplied by consumer $p_j$, $s^{p_j}_t$ is the power allocated to it and $\alpha^{p_j}_t$ parameterizes its pre-announced PQ graph function $q^{p_j}_{\alpha^{p_j}_t}(\cdot)$. Finally, the prosumers submit their $N_c+N_p$ PQ graph parameters to the market clearing price functional $f^M(\cdot)$ that determines the clearing price $\alpha_t.$ The system dynamics are given by
\begin{equation}
    \begin{aligned}
    d^i_{t+1} &= f^{c_i}\left( d^i_{t},\alpha_t,q^{c_i}_{\alpha^{c_i}_t} \right) + n^{c_i}_t \\
    \alpha^{c_i}_{t+1} &= \alpha^{c_i}_t + u^{c_i}_t, i\in[N_c] \\
    s^j_{t+1} &= f^{p_j}\left( s^j_{t},\alpha_t,q^{p_j}_{\alpha^{p_j}_t} \right) + n^{p_i}_t \\
    \alpha^{p_j}_{t+1} &= \alpha^{p_j}_t + u^{p_j}_t, j\in[N_p] \\
    \alpha_t &= f^M\left(\left(q^{c_i}_{(\cdot)}\right)_{i\in[N_c]},\left(q^{p_j}_{(\cdot)}\right)_{j\in[N_p]}\right),
    \end{aligned}
    \label{eqn:priceAnticipatingSyetemModelGENERAL}
\end{equation}
where $\left\lbrace n^{c_i}_t\right\rbrace_{\geq0}$ and $\left\lbrace n^{p_j}_t\right\rbrace_{\geq0}$ are the random state noise processes. In \eqref{eqn:priceAnticipatingSyetemModelGENERAL}, the strategic actions of the consumers and suppliers at time $t$ are denoted by $u^{c_i}_t$ and $u^{p_j}_t$ respectively.  
The functions $f^{c_i}$ and $f^{p_j}$ are, respectively, assumed to be strictly decreasing and strictly increasing in $\alpha_t.$ Next, we adopt a \emph{Linear Economy} model \cite{acemoglu-etal17competition-electricity-markets-renewable} with PQ graphs of the form $q^{c_i}_{\alpha^{c_i}}(p):=\alpha^{c_i}-p,~\forall i\in[N_c]$ and $q^{p_j}_{\alpha^{p_j}}(p):=\alpha^{p_j}+p,~\forall j\in[N_p]$,
  and assume that the clearing price functional is of the form $f^M\left(\left(q^{c_i}_{(\cdot)}\right)_{i\in[N_c]},\left(q^{p_j}_{(\cdot)}\right)_{j\in[N_p]}\right) := \frac{\kappa}{N_c+N_p}\left(\sum_{i\in[N_c]}\alpha^{c_i}+\sum_{j\in[N_p]}\alpha^{p_j}+\zeta\right),$ for some $\kappa,\zeta\in\mathbb{R}.$

Standard market models define a utility function for consumers that includes a convex \emph{blackout cost} and profit functions for the suppliers both of which inform the optimization problem that the RTO then solves. However, following the development in \cite[Sec.~II]{kizilkale-mannor10volatility-efficiency-markets-friction}, albeit in discrete time, we approximate both these functions with costs that are \emph{quadratic} in the state. Moreover, the quadratic approximation is commonly employed in the literature because it constitutes a second-order \enquote{least-squares} approximation to other nonlinear cost functions\footnote{For example, production cost and blackout costs are modeled as convex functions and hence, can be well approximated with quadratic functions.} and because it is analytically tractable \cite{basar-olsder98dynamic--noncoop-game-theory}, admitting closed-form equilibrium solutions that as we show, provide interesting insights into market behavior. Consequently, each market participant has a cost function it needs to minimize given by
\begin{eqnarray}
J^{(r)}_{c}\left(\xc_0\right) & := & \frac{1}{2}\E\sum_{t=0}^{\infty}\gamma^t\left({\xc}^TQ_i\xc_t + r (u^{c_i}_t)^2\right),~i\in[N_c],\text{ and } \label{eqn:consumerCostFunction}\\
J^{(r)}_{p}\left(\xp_0\right) & := & \frac{1}{2}\E\sum_{t=0}^{\infty}\gamma^t\left({\xp}^TQ_j\xp_t + r (u^{p_j}_t)^2\right),~j\in[N_p], \label{eqn:producerCostFunction}
\end{eqnarray}
where, as before, $\left(Q_i\right)_{i=1}^{N_c}$ and $\left(Q_i\right)_{j=1}^{N_p}$ are assumed positive semidefinite, and the \emph{control penalty coefficient} $r>0.$ 

Finally, \cite{kizilkale-mannor10volatility-efficiency-markets-friction} further assumes that the demand and production processes are linear mean-reverting (i.e., Ornstein–Uhlenbeck) processes, and their time discretization results in the following model for the market participants.
\begin{eqnarray}
  \xc_{t+1} & = & A^{c_i}\xc_t + \bc u^{c_i}_t + \nos^{c_i}_{t},~i\in[N_c],\label{eqn:consumerStateEvolution}\\
  \xp_{t+1} & = & A^{p_j}\xp_t + \bp u^{p_j}_t+ \nos^{p_j}_{t},~j\in[N_p],\text{ and }\label{eqn:producerStateEvolution}\\
  \alpha_t & = & \frac{\kappa}{N_c+N_p}\left(\sum_{i\in[N_c]}\alpha^{c_i}+\sum_{j\in[N_p]}\alpha^{p_j}+\zeta\right),\nonumber
\end{eqnarray}
where, the matrices $A^{c_i},\bc$ are of the form 
\begin{equation*}
A^{c_i} = 
    \begin{bmatrix}
    \ast  &  0  &  \ast      \\
    0  & \ast & \ast \\
    0 & 0 & 0
\end{bmatrix},
\bc = 
\begin{bmatrix}
    0 \\
    0 \\
    1
\end{bmatrix},
\end{equation*}
and $A^{p_j}, \bp$ are of the form
\begin{equation*}
A^{p_j} = 
    \begin{bmatrix}
    \ast  &  \ast      \\
    0 & 0 
\end{bmatrix},
\bp = 
\begin{bmatrix}
    0 \\
    1
\end{bmatrix}.
\end{equation*}
Note that the dynamics of the prosumers are all coupled through the clearing price process $(\alpha_t)_{t\geq0}.$
Let $N_{\mathcal{M}}:=3N_c+2N_p$. The state vectors can now be aggregated into a single vector $\x:=[(\xc)_{i\in[N_c]};(\xp)_{i\in[N_p]}]\in\mathbb{R}^{N_{\mathcal{M}}}$, and a giant coefficient matrix $A\in\mathbb{R}^{N_{\mathcal{M}}}\times\mathbb{R}^{N_{\mathcal{M}}}$ suitably defined using $(A^{c_i})_{i\in[N_c]},(A^{p_j})_{j\in[N_p]}$, and $\bc$ and $\bp$ redefined to vectors in $\mathbb{R}^{N_{\mathcal{M}}}$. This aggregate state then evolves as follows,
\begin{eqnarray}
  \x_{t+1} = A\x_t + \sum_{i=1}^{N_c}\bc u^{c_i}_t + \sum_{j=1}^{N_p}\bp u^{p_j}_t + \mathbf{n}_t,~t\geq0.
  \label{eqn:priceAnticipatingFULLMarketStateEvolution}
\end{eqnarray}
and will be the object of our study in the sequel. Note, once again, that the decisions $u^{c_i}_t$ and $u^{p_j}_t$ in $\x_t$ are taken in a \emph{decentralized} manner, the price process $\alpha_t$ is set in a \emph{centralized} manner and causes all prosumer dynamics to be coupled.

\subsection{Existence and Properties of the Markov Perfect Equilibrium}\label{sec:markovPerfectEqubmProperties}
While the analysis presented below works with any number $N_c,N_p$ of consumers and producers, we restrict ourselves to a single producer and consumer, i.e., $N_c = N_p = 1$ for simplicity. Alternatively, this can be viewed as the aggregate of a number of infinitesimal statistically identical producers and consumers. Further, since the system is fully observed and linear quadratic, we will prove all results for the noiseless case (i.e., ignoring $\mathbf{n}_t$ in \eqref{eqn:priceAnticipatingFULLMarketStateEvolution}) and appeal to certainty equivalence \cite{bertsekas95dynamic-programming-optimal-control-volume1} where necessary without loss of any generality. In this section, we first show that the system in \eqref{eqn:priceAnticipatingFULLMarketStateEvolution} has a Nash Equilibrium strategy profile and derive some important properties.

Before we state the central result of this section, a few definitions are in order. First, following standard game theoretic convention, $-i$ will refer to Player~$2$ when $i=1,$ and vice versa. Next, consider the (stationary) control policy ${\pi^*}^i:=[{\mu^*}^i,{\mu^*}^i,\cdots],~i\in\{1,2\},$ defined as 
\begin{eqnarray}
  {u^*}^i_t = {\mu^*}^i(\x_t) := -\p_i^T\x_t,~t\geq0,
  \label{eqn:stationaryNashFeedbackControlPolicy}
\end{eqnarray}
where the ${\p}_i,i\in\{1,2\}$ satisfy the following recursion
\begin{equation}
\begin{aligned}
\gamma (\b^i)^T K^i A & = \left(r + \gamma(\b^i)^T K^i \b^i \right)\mathbf{p_i}^T + \gamma(\b^i)^T K^i \b^{-i}\p_{-i}^T,\\
K^i & =  \gamma F^T K^i F + r \p_i\p_i^T + Q_i ,\\
F & :=  A - \sum_{i=1}^2\b^i\p_i^T.\label{eqn:riccattiNashRecursion}
\end{aligned}
\end{equation}
Also define matrices $F_i,~i\in[2]$ by
\begin{equation*}
F_i = F + \b^i\p_i^T.
\end{equation*}
\begin{asmpn}\label{assmpn:nashControllabiilityObservability}
There exist vectors $\p_i$ and matrices $K^i,~i\in\{1,2\}$ satisfying \eqref{eqn:riccattiNashRecursion}. Furthermore, the pairs $(F_i,\b^i)$ are controllable and $(F_i,Q_i)$ are observable.
\end{asmpn}

\begin{thm}[\cite{basar-olsder98dynamic--noncoop-game-theory}]
Under Assumption~\ref{assmpn:nashControllabiilityObservability}, the dynamics described in \eqref{eqn:priceAnticipatingFULLMarketStateEvolution} together with the cost functions \eqref{eqn:consumerCostFunction} and \eqref{eqn:producerCostFunction} admits a Nash equilibrium strategy profile given by $\pi^*:=[{\pi^*}^1,{\pi^*}^2]$ as defined in \eqref{eqn:stationaryNashFeedbackControlPolicy} that leads to the equilibrium cost function 
\begin{eqnarray}
  J^*_i(\x_0,r) := \frac{1}{2}{\x_0}^T K_i \x_0,~i\in[2].
  \label{eqn:nashEqCostFunction}
\end{eqnarray}
\end{thm}
Moreover the resulting dynamics, described by $\x_{t+1}=F\x_t,t\geq0,$ are stable. 

\subsubsection{Time Consistency and Subgame Perfection}
We now investigate some properties of the above equilibrium profile.
\begin{defn}[Markov Perfection, \cite{maskin-tirole01markov-perfect-equilibrium}]
A Nash equilibrium strategy profile is said to be \emph{\color{blue}Stationary Markov Perfect} if it is
\begin{itemize}
    \item \emph{Markovian,} i.e., each player's strategy depends only on the current state of the game,
    \item \emph{Stationary,} i.e., each player's strategy at every time $t$ depends only on the current state $\x_{t}$ and 
    \item \emph{Subgame Perfect.}
\end{itemize}
\end{defn}
While the first two points are quite evident, the last can be argued by noting the fact that the equilibrium strategy \eqref{eqn:stationaryNashFeedbackControlPolicy}, \emph{\color{blue}Strongly Time Consistent} (see \cite[Thm.~6.10]{basar-olsder98dynamic--noncoop-game-theory}). We will now define Time Consistency and explain why Strong Time Consistency implies Subgame Perfection. Let the tuple $D(\Pi,\mathsf{N})$ denote the $N$-person dynamic game whose Nash equilibrium is being studied, where $\Pi:=\bigtimes_{i=1}^N\Pi_i$ is the product strategy space and $\mathsf{N}=\{0,1,\cdots\}$ is the decision period. Further, let $\pi_{[s,t]}\in\Pi_{[s,t]}$ and $\pi_{[s,t]}\in\Pi_{[s,t]}$ denote the truncations of $\pi\in\Pi$ and $\pi_i\in\Pi_i$ to the interval $\{s,s+1,\cdots,t\}$, with $[t,\infty]:=\{t,t+1,\cdots\},$ and define 
\begin{equation}
    D^\beta_{s}:=D\left(\left\lbrace\pi\in\Pi:\pi_{[0,s-1]}=\beta_{[0,s-1]} \text{ and } \pi_{[s,\infty]}\in\Pi_{[s,\infty]}\right\rbrace , \mathsf{N} \right),
    \label{eqn:subgameDefinition}
\end{equation}
to be the version of the game $D(\Pi,\mathsf{N})$ where the policies of all players over $[0,s-1]$ are fixed as $\beta^i_{[0,s-1]}\in\Pi_{[0,s-1]},~i\in[N]$.

\begin{defn}[Time Consistency]
\begin{enumerate}
    \item An $N$-tuple of policies $\pi^*\in\Pi$ solving $D(\Pi,\mathsf{N})$ is said to be \emph{Strongly Time Consistent} (STC) if its truncation to $[t,\infty]$, $\pi^*_{[t,\infty]}$, solves the truncated game $D^\beta_t,~\forall\beta_{[0,t]}\in\Pi_{[0,t]}$ and $t\in\mathsf{N}.$
    \item An $N$-tuple of policies $\pi^*\in\Pi$ solving $D(\Pi,\mathsf{N})$ is said to be \emph{Weakly Time Consistent} (WTC) if its truncation to $[t,\infty]$, $\pi^*_{[t,\infty]}$, solves the truncated game $D^{\pi^*}_t,~\forall t\in\mathsf{N}.$
\end{enumerate}
\end{defn}
By definition, we see that a policy that is STC is always Subgame Perfect, which concludes our proof of Markov Perfection of the strategy profile \eqref{eqn:stationaryNashFeedbackControlPolicy}. However, it is to be noted that Weak Time Consistency does not imply Subgame Perfection, as demonstrated in \cite{fershtman89time-consistency-subgame-perfection}. We now move on to proving other properties of the equilibrium solution. 

\subsubsection{Bellman Best Response Correspondence}
As in the centralized control case, the Equilibrium cost function, $J^*_i(\cdot,\cdot)$ satisfies a Bellman Best Response Correspondence equation
\begin{eqnarray}
  J^*_i(\x,r) = \x^T Q_i \x + \inf_{u\in\mathbb{R}} r~u^2 + \gamma J^*_i(A\x+\b^i u+ \b^{-i}{u^{-i}}^*,r).
  \label{eqn:bellman4Nash}
\end{eqnarray}
 The proof follows from recognizing the fact that the optimal cost is of the form \eqref{eqn:nashEqCostFunction}, and using  \eqref{eqn:riccattiNashRecursion} to establish an identity. Note that in \eqref{eqn:bellman4Nash}, the strategy of the other player(s) is held at Nash. 
Finally, as in Sec.~\ref{sec:lqrModelPreliminaries}, we call a control policy $\pi\in\Pi$ \emph{\color{blue}admissible} if it the cost and volatility functionals are well defined under $\pi$ and denote by $\mathcal{A}\subset\Pi$ the subset of all admissible policies. We define a \emph{\color{blue}Bellman Operator} as follows. Let $\mathcal{S} = \reals^d\times(0,\infty)$ and let $\mathcal{J} = \left\lbrace v\mid v:\mathcal{S}\rightarrow \mathbb{R}\right\rbrace$ be the space of all functionals on $\mathcal{S}$. The Bellman Operator $T_i:\mathcal{J}\rightarrow\mathcal{J}$ is given by
\begin{equation}\label{eqn:Bellman4NashOperator}
    (T_iv)(\x,r) := \x^TQ_i\x + \inf_{u\in\mathbb{R}} r~u^2 + \gamma v(A\x+\b_i u+ \b_{-i}u^*_{-i},r),~i\in[2],
\end{equation}
and \eqref{eqn:bellman4Nash} shows that $J^*_i$ is a fixed point of this operator. Now that the requisite operator has been defined, we will invoke arguments akin to those in Sec.~\ref{sec:regulatorEffVersusControlVolatilityinLQR} to show the volatility-efficiency tradeoff.

\subsection{Volatility-Efficiency Tradeoff}
Define a per-prosumer state penalizing cost by
\begin{eqnarray}
  J^{(i)}_{sp,r}(\x_0) := \sum_{t=0}^\infty \gamma^t\x^T_tQ\x_t = -\mathcal{E}_{\x_0}(\pi^*),
\label{eqn:gammaDiscountedCostNASHLinearQuadraticSystemNoControlPenalty}
\end{eqnarray}
where the subscript $sp$ stands for \enquote{state penalizing.} This is the cost of applying ${\pi^*}^i$ to the system
\begin{eqnarray}
  \x_{t+1} & = & \tilde{A}_i\x_t + \b_iu^i_t,~t\geq0,\text{ where }\nonumber\\
  \tilde{A}_i & := & A - \b^{-i}p_{-i}.
\end{eqnarray}
Moreover, the \emph{\color{blue}Social cost} of the Nash Equilibrium strategy is defined\footnote{The \enquote{\color{blue}$\mathcal{N}$} in the superscript indicates that this is computed at Nash.} as $J^{\mathcal{N}}_{sp,r}(\x_0) := \sum_{i=1}^2J^{(i)}_{sp,r}(\x_0).$ 
Arguing along the same lines as Thm.~\ref{thm:CostWithOutControlPenaltyConcaveInR}, we see that $J^{(i)}_{sp,r}$ is concave, nondecreasing in $r,$ whereby
\begin{tcolorbox}

\begin{thm}[V-E tradeoff]\label{thm:nashEqb,CostWithOutControlPenaltyConcaveInR}
The social cost function $J^{\mathcal{N}}_{sp,r}(\x)$ is concave nondecreasing in $r.$ Specifically,
\begin{equation}
\begin{aligned}
\frac{dJ^{\mathcal{N}}_{sp,r}(\x)}{dr} &> 0,\text{ and }\\
\frac{d^2J^{\mathcal{N}}_{sp,r}(\x)}{dr^2} &\leq 0,~\forall \x\in\mathbb{R}^{N_{\mathcal{M}}}.\label{eqn:nashEqbmCostWithoutControlPenalty1stAnd2ndDerivatives}
\end{aligned}
\end{equation}
\end{thm}

\end{tcolorbox}

Thus, we see that the equilibrium strategy, while possessing several desirable characteristics such as being Nash-Walras and Markov perfect, results in an inevitable tradeoff between social cost and price volatility. In anticipation of concerns that the strategic or competitive behavior of market participants might be at the root of this phenomenon, we next show that even with passive prosumers, this market behavior cannot be avoided.


\section{Applications to the Analysis of Deregulated Electricity Markets, Part II: Price Taking Participants}\label{sec:applicationPriceTakingMarketAnalysis}
In this section, we drop the price-anticipating assumption of prosumers and focus on a market with price-taking participants, which means that the suppliers and consumers are no longer strategic. Linear dynamical system models have often been used to model markets with friction in general and electricity markets in particular in the literature \cite{alvarado97dynamics-power-markets,alvarado99stability,watts-alvarado04influence-futures-markets-price-stabilization-electricity,nutaro-protoSomething09impact-signal-delay-stability-power-markets}, including the \emph{linear quadratic cost} model \cite{mcnelis-asilis02macroeconomic-games-asset-volatility-lqr-france-germany,okajima2013dynamic-mechanism-linear-quadratic-pricing-delay,kizilkale-etal10regulation-efficiency-markets-friction,berger-schweppe89real-time-pricing-load-frequency-control,kizilkalemannor11regulation-double-price-mechanisms}. The state of the system under study at time $t\in\mathsf{N} = \{0,1,\cdots\}$, is described by the (demand, supply, price) tuple denoted by $\x_t=[d_t,s_t,p_t]^T$ and evolves in $\mathbb{R}^3$ according to the following law (akin to Eqn.~\eqref{eqn:LinearSystemWithConstForcingFunction})
\begin{equation}
\x_{t+1} = A\x_{t}+\b u_t+\mathbf{n}_t,t\in\mathsf{N}. 
\label{eqn:linearPowerMarket}
\end{equation}

As is quite common in economic literature \cite{cho-meyn10efficiency-marginal-cost-dynamic-competitive-markets-friction,kizilkale-etal10regulation-efficiency-markets-friction}, the quantities $d_t$ and $s_t$ represent the aggregate of a continuum of infinitesimally small, identical consumers and energy suppliers. In the sequel, we deal with aggregate demand and supply. Taking into consideration standard facts in microeconomic theory that demand decreases with price and supply increases with price \cite{mas-collel-etal95microeconomic-theory}, the matrix $A$ is of the form ($\beta,\sigma,\phi_1,\phi_2$ below represent nonnegative reals)
\begin{eqnarray}
A =\left[ \begin{matrix}
  \beta & 0 & -\phi_1 \\
  0 & \sigma & \phi_2 \\
  0 & 0 & 1
 \end{matrix}\right],
 \label{eqn:structureOfMatrixA}
\end{eqnarray}
and $\b = [0, 0, 1]^T$. The regulatory authority (ISO or RTO) controls the price of electricity by tweaking the control input $u_t$ in each time slot and the suppliers and consumers are coupled through this price. Therefore, $u_t$ controls the rate of change of the price of electricity. Recall that the ISO or the RTO is only allowed to control the system through the \emph{price} signal. This constraint is reflected in the value of $\b = [0, 0, 1]^T.$ In the sequel, we will assume that the matrix $A$ along with disturbance covariance matrix  $\boldsymbol{\Sigma}:=\mathbb{E}\mathbf{n}_t\mathbf{n}^T_t$ are known to the RTO/ISO.

The definitions of the cost, volatility and efficiency functionals remain unchanged, but we will refer to them in this section and Sec.~\ref{sec:simulationResults} as {\color{blue}Social Cost}, {\color{blue}Price Volatility} and {\color{blue}Market Efficiency} respectively.  Market Efficiency ($\mathcal{E}_{\x_0}(\cdot)$) is a notion of social welfare. This, therefore, is something the regulatory authority (RTO/ISO) must attempt to maximize using a suitable pricing mechanism. 

Before we discuss optimal pricing policies, we note that Observability is easily guaranteed by choosing $Q$ to be positive definite, which can be ensured by minimally perturbing the chosen $Q$ with a scaled identity matrix. The resulting quadratic cost functional, given, as in Sec.~\ref{sec:lqrModelPreliminaries}, by
\begin{eqnarray}
    J_r(\x_0) = \mathbb{E}\sum_{t=0}^\infty \gamma^t \left(\x^T_tQ\x_t + r~u^2_t\right),
              \label{eqn:gammaDiscountedCostLQRMarket}
\end{eqnarray}
has been employed in the literature as a differentiable (and hence, more tractable) approximation to other more general cost functions (see for example, \cite[Sec.~IV]{kizilkale-mannor10regulation-efficiency-markets-friction}). To guarantee Controllability, requires the matrix $C_0=[b~Ab~A^2b]$ is required to be full rank. In our case, 
\begin{eqnarray*}
C_0 =\left[ \begin{matrix}
  0 & \phi_1 & \phi_1 (\beta+1)\\
  0 & -\phi_2 & -\phi_2 (\sigma+1)\\
  1 & 1 & 1
 \end{matrix}\right].
\end{eqnarray*}
This matrix is full rank when $\beta\neq\sigma,$ which is the case in practice, since demand and supply rarely decrease or increase at exactly the same rate. 

\subsection{Price Volatility and Market Efficiency}\label{sec:priceVoltyMktEffcy}
To begin with, consider the implications of Thm.~\ref{thm:CostWithOutControlPenaltyConcaveInR} in the present context. With regards to deregulated markets, the result essentially means that efficiency increases with increasing volatility, establishing a clear volatility-efficiency tradeoff. We thus see that operating an electricity market at high efficiency automatically entails highly volatile prices. The simulation results presented in Sec.~\ref{sec:simulationResults} illustrate this phenomenon very clearly. So, the aforementioned volatility in market prices is actually a \emph{natural outcome} of attempting to improve market efficiency.

Moving on to Thm.~\ref{thm:CapacityRegionOfMarket}, we see that with every Market~$(A,\b,F_n,Q,r,\gamma),$ is associated a region $\mathcal{C}\subset\mathbb{R}^2$ such that all achievable $(Volatility,~Efficiency)$ pairs lie within $\mathcal{C}$ and, in keeping with our nomenclature, we term this set the {\color{blue}Capacity Region} of Market~$(A,\b,F_n,Q,r,\gamma).$ This latter observation is of practical importance since it can be used to inform RTOs and ISOs about the consequences of price control policy decisions on market efficiency. Pricing policies are the outcome of decisions made within federal and local governments, and this analysis helps quantify the social impact such decisions entail. In general, these consequences are not very easy to judge off-hand. One recent example is the enormous price spike experienced by ERCOT where price capping and scarcity pricing policies lead to electricity being priced at $\$9000$/MWh and yet, defenders of the policy cited it as \enquote{a more efficient system} \cite{daprato19texas-power-spike-carbon-free}. 
 

\subsection{Effects of Decarbonization}\label{sec:renewablesAndTheVolatilityCliff}
In recent times, rejecting an incremental approach to solving climate  change, undertakings such as the Deep Decarbonization Pathways Project (DDP) have focused on introducing drastic reductions in dependence on carbon-based fuel sources \cite{wiki19deep-carbonization-pathways-project, pye16improving-deep-carbonization-modelling}. In Sec.~\ref{sec:Introduction}, we outlined the deleterious effects that the entry of renewable energy sources can have on electricity prices. We now quantify the effects of endeavours such as the DDP on existing electricity markets using a modified system model, to incorporate these intermittent power sources and, in doing so, uncover a phenomenon we call a \enquote{volatility cliff.} 

The output of a renewable energy source is highly dependent on ambient weather conditions. Solar power generation (even with controllable solar panels) is almost entirely at the mercy of the sun and local cloud cover. Similarly, windmills produce more power when the weather is windy. 
Taking this unique difficulty in controlling the output of these sources into consideration, we model renewable sources in the supply as a controlled random walk as follows
\begin{equation}\label{eqn:renewableSupplyStateEvolution}
y_{t+1} = \sigma_r y_t + \sigma_c p_t + w_t,~t\geq0.
\end{equation}
Here, at time $t$, $y_t$ is the aggregate output of installed renewable sources, $p_t$ is the price control signal, $w_t$ is the state disturbance and $\sigma_r$ and $\sigma_c$ are a positive constants. The price coefficient $\sigma_c$ is a small positive constant used to model whatever rudimentary control the price signal can have over renewable supply. To incorporate this new evolution and keep the controlled process Markovian, we define the new system state vector to be $\mathbf{z}_t = [d_t,s_t,p_t,y_t]^T,t\geq0.$ We note that the state now evolves in $\mathbb{R}^4$ as opposed to the state in \eqref{eqn:LinearSystemWithConstForcingFunction} that evolved in $\mathbb{R}^3.$ Therefore, the evolution equation \eqref{eqn:LinearSystemWithConstForcingFunction} is modified to
\begin{equation}
  \mathbf{z}_{t+1} = A_r\mathbf{z}_t + \b_r u_t + \mathbf{n}^{(r)}_t,t\geq0,
  \tag{REN}
  \label{eqn:renewablesStateEvolution}   
\end{equation}
where 
\begin{eqnarray*}
A_r =\left[ \begin{matrix}
  A & \mathbf{a}^{(1)}_r\\
  \left(\mathbf{a}^{(2)}_r\right)^T & \sigma_r
 \end{matrix}\right]_{4\times4},
\end{eqnarray*}
$\b_r = [0, 0, 1, 0]^T$, and the state disturbance $\mathbf{n}^{(r)}_t = [\mathbf{n}_t w_t]^T$ consists of IID random vectors with $\mathbb{E}\mathbf{n}^{(r)}_t = \mathbf{0}$ and  $\mathbb{E}\mathbf{n}^{(r)}_t(\mathbf{n}^{(r)}_t)^T = \Psi^{(r)}_n.$ The vector $\mathbf{a}^{(1)}_r = [0, 1, 0]^T$, $\mathbf{a}^{(2)}_r = [0, 0, \sigma_c]^T$ and, assuming the state disturbance for supply, demand and renewables are independent of each other, the covariance matrix $\Psi^{(r)}_n$ is a diagonal matrix, with diagonal entries\footnote{We assume the price process is completely under the control of the regulatory authority and hence, doesn't have any noise in its evolution.} $[\psi_d,\psi_s,0,\psi_r]$. Single stage cost is defined, as before, to be $g(\mathbf{z}_t,u_t) = \mathbf{z}^T_tQ\mathbf{z}+r~u^2_t,~t\geq0,$ and the cost functional $J^{(ren)}_r$ and efficiency $\mathcal{E}^{(ren)}$ are both defined as before, replacing $\x$ with $\mathbf{z}$. 

As renewable supplies proliferate (a) the aggregate supply from these sources obviously increases and (b) supplied power becomes more volatile. The factor in Eqn.~\eqref{eqn:renewablesStateEvolution} that captures both of these effects is the renewable supply variance $\psi_r.$ As the contribution of renewables increases, so does $\psi_r.$ We then have the following result for the system \eqref{eqn:renewablesStateEvolution}.
 
 \begin{prop}\label{prop:moreRenewablesMorePriceVolatility}
 Consider two renewable sources with supply variance $\psi^{(1)}_r$ and $\psi^{(2)}_r$. Also denote the price volatility of the system using renewable Source~$i$ by $\mathcal{V}^*_i,~i=1,2.$ Then, if $\psi^{(1)}_r < \psi^{(2)}_r,$  $\mathcal{V}^*_1 < \mathcal{V}^*_2.$
 \end{prop}
 
 \begin{pf}
 See Sec.~\ref{appendix:proofOfPropmoreRenewablesMorePriceVolatility} in the Appendix.
 \end{pf}

Prop.~\ref{prop:moreRenewablesMorePriceVolatility} leads us to conclude that as renewable energy sources become more widespread in the power grid, increased variability of prices in the \emph{current} market structure is inevitable. This result then, buttresses the arguments presented in recent studies such as \cite{joskow19challenges-wholesale-intermittent-renewable-at-scale} that call for a restructuring of the electricity market including and especially the pricing pricing process. 

In fact, as the result of simulating this system shows in Fig.~\ref{fig:capacityRegionShrinksWithIncreasingRenewables}, the effect is even more pernicious, in that increasing the fraction of renewable supply in the market causes the entire market capacity region to shrink. However, we reiterate that the results in this and the next section are not intended to stymie efforts towards decarbonization. Rather, they seek to highlight how the current market model might be insufficient to handle emerging green technologies such as wind and photovoltaic power supplies.

\section{Numerical Results}\label{sec:simulationResults}
We now discuss the results of various simulation experiments conducted to support the theory we have developed in earlier sections. 
\subsection{The Volatility-Efficiency tradeoff in LQRs}
\begin{figure}[tbh]
\centering
\includegraphics[height=5.70cm, width=11.50cm]{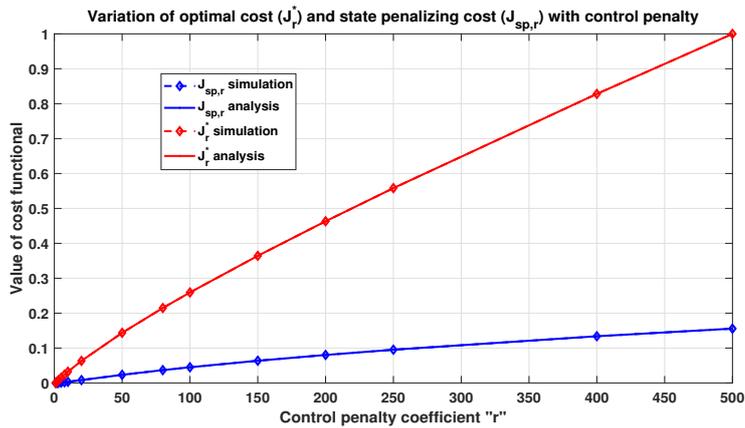}
\caption{Variation of the two cost functionals the optimal cost $J^*_r$ and the state penalizing cost $J_{sp,r}$ as functions of the control penalty $r$ under the optimal pricing policy $\pi^*$ in \eqref{eqn:optimalCENTRALIZEDControlLQGSystem}. In this simulation, $\x_t\in\reals^8.$ The plot shows that both $J^*_r$ and $J_{sp,r}$ are concave and increasing in $r.$ Note that both curves have been normalized to lie in $[0,1].$ }
\label{fig:optimalAndStatePenalizingCostsConcaveInR}
\end{figure}

We begin with simulating the Linear Quadratic Regulator in \eqref{eqn:LinearSystemWithConstForcingFunction}. The LQR evolves in $\reals^8,$ and the entries of the system parameters $(A_{8\times8},\b_{8\times1},Q_{8\times8})$ are sampled according to the standard normal distribution. The state disturbance process is assumed to be IID Gaussian, i.e., $F_n\equiv\mathcal{N}(\mathbf{0},I_{8\times8})$, and the discount factor $\gamma = 0.5.$
The simulation results shown in Fig.~\ref{fig:optimalAndStatePenalizingCostsConcaveInR} help corroborate Thm.~\ref{thm:CostWithControlPenaltyConcaveInR} and Thm.~\ref{thm:CostWithOutControlPenaltyConcaveInR}. Because the control input to the system is implicitly constrained while using large values of $r$, even though the control signal shows little volatility, the state variables show larger fluctuations, increasing the \emph{state penalizing} portion in every time slot (Fig.~\ref{fig:2MarketSamplePathsWithControlPenalty} and its accompanying explanation illustrate this phenomenon quite clearly). Furthermore, comparing Equations \eqref{eqn:OptimalGammaDiscountedCostCENTRALIZEDLinearQuadraticSystemWithControlPenalty} and \eqref{eqn:gammaDiscountedCostCENTRALIZEDLinearQuadraticSystemNoControlPenalty} we see that the optimal cost function $J^*_r(\cdot)$ includes a non negative control penalty term in addition to the state penalty, which makes it the larger of the two metrics. As can be seen in the figure, both $J^*_r(\cdot)$ and the state penalizing cost function $J_{sp,r}(\cdot)$ are increasing and concave in the control penalty coefficient $r.$

\begin{figure}[tbh]
\hspace{-0.5cm}%
\centering
\includegraphics[height=5.7cm, width=12.00cm]{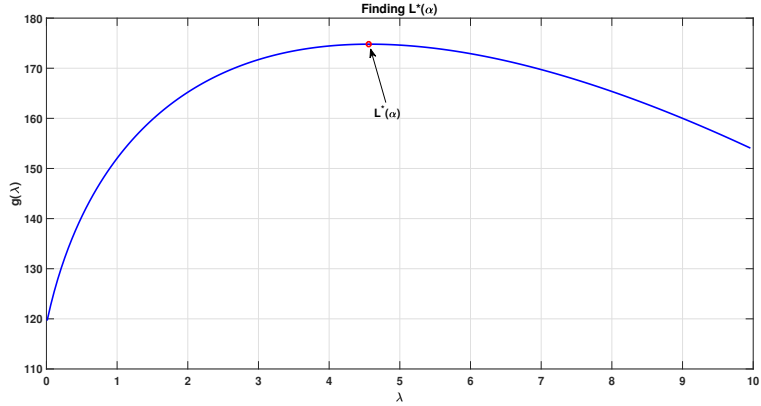} 
\caption{Illustrating the behavior of the function $q_\alpha$, defined in Eqn.~\eqref{eqn:gOfLambda}, for the market whose parameters are given in Sec.~\ref{app:detailsOfeExperiments}, at $\alpha=27$. The function, as argued in Sec.~\ref{sec:CapacityRegionNoFreeLunchTheorem}, is \emph{concave} in $\lambda$. The value of the constrained MDP for a given acceptable volatility level $\alpha$, $L^*(\alpha)=\sup_{\lambda\geq0}q_\alpha(\lambda)$, is also shown in the figure.}
\label{fig:gOfAlphaIsConcave}
\end{figure}

\begin{figure}[tbh]
\centering
\includegraphics[height=5.7cm, width=12.00cm] {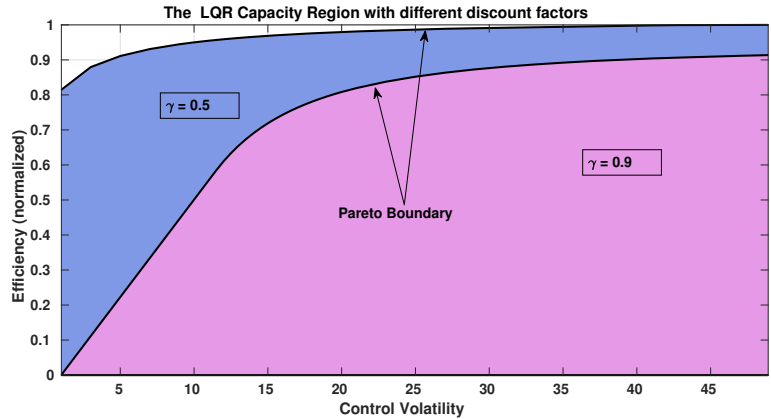}
\caption{The capacity regions with discount factor $\gamma = 0.5$ and $\gamma = 0.9$ (\emph{normalized} to peak at 1 with $\gamma = 0.5$) for the LQRs whose parameters $(A,F_n,Q)$ are given in  Sec.~\ref{app:detailsOfeExperiments}. The control penalty is set to $r = 0.01.$ The Pareto boundary is defined by the optimal volatility function $\mathcal{E}^*$. The regions below these boundaries, shaded blue and pink, are the capacity regions with $\gamma = 0.5$ and $\gamma = 0.9$ respectively.}
\label{fig:marketCapacityRegionSimulated}
\end{figure}

Moving on to LQR capacity regions, for a given $(A,\b,F_n,Q,r,\gamma)$, define for every price volatility level $\alpha,$ the function
\begin{equation}
q_\alpha(\lambda) = \x_0K_\lambda\x_0 + \frac{\gamma}{1-\gamma}\mathbb{E}\mathbf{n}^TK_\lambda\mathbf{n}-\lambda \alpha.
\label{eqn:gOfLambda}
\end{equation}
We see that the optimal value of the constrained MDP $\mathcal{L}^* = \sup_{\lambda\geq0}q_\alpha(\lambda)$ (see Eqns.~\eqref{eqn:LagrangianCMDP4VETradeoff} and \eqref{eqn:optimalCostOfConstrainedMDP}). The arguments presented in Sec.~\ref{sec:CapacityRegionNoFreeLunchTheorem} and in Thm.~\ref{thm:CapacityRegionOfMarket} show that both $q_\alpha()$ and $\mathcal{L}^*$ are concave functions of $\lambda$ and $\alpha$ respectively. Figures~\ref{fig:gOfAlphaIsConcave} and \ref{fig:marketCapacityRegionSimulated} show that this is indeed true. 

Fig.~\ref{fig:marketCapacityRegionSimulated} shows the capacity region associated with \emph{two} LQRs with $\gamma = 0.5$ and $\gamma = 0.9$ respectively. For a given value of $\alpha,$ suppose that the optimal control with $\gamma = 0.9$ is $\{u^{(0.9)}_t,t\geq0\}$ and $\mathcal{L}^*_{0.9}$ is the optimal cost. When this same control is used in a market with $\gamma = 0.5$ starting with the same initial conditions, the resulting sample paths are identical, but the cost only reduces, since $\gamma$ has reduced. But since applying the optimal control $\{ u^{(0.5)}_t,t\geq0\}$ can only further reduce cost, the efficiency with the latter discount factor must be larger, giving rise to a bigger capacity region. We see Fig.~\ref{fig:marketCapacityRegionSimulated} bearing this out.
Note that in Fig.~ \ref{fig:marketCapacityRegionSimulated}, the maximum attainable efficiency curve $\mathcal{E}^*_{\x_0}$ has been normalized to peak at $1$ for $\gamma = 0.5.$ We also refer to this curve as the \enquote{Pareto Boundary} since no achievable vector can give strictly greater efficiency for a given level of volatility. The region below the $\mathcal{E}^*_{\x_0}$ curve, shaded blue and pink, are the capacity regions of the markets, i.e., every $(\mathcal{V},\mathcal{E})$ tuple in these regions is achievable by some admissible policy.

\subsection{Market with Price-anticipating Prosumers}
Next, we move on to a more concrete example of a market. 
Fig.~\ref{fig:2MarketSamplePathsWithControlPenalty} shows the effect of the control penalty coefficient $r$ in \eqref{eqn:quadraticCostWithControlPenalty}. Four sample paths of the of the market equilibrium price process $\{\alpha_t,t\geq0\}$ are shown demonstrating the effects of light control penalty ($r=0.1$) to heavy control penalty ($r=100$). In all cases, we begin with the same initial condition, but owing to the effect of the control penalty, the sample paths differ significantly in their evolution, with the price fluctuations being considerable with $r=0.1.$ However, bearing out  Thm.~\ref{thm:nashEqb,CostWithOutControlPenaltyConcaveInR}, we see that $J^{\mathcal{N}}_{sp,r}= 9.552\times10^7\text{ and }1.468\times10^8$ with $r=0.1$ and 100, resp, clearly demonstrating the Volatility-Efficiency tradeoff at the Nash-Walras equilibrium strategy profile.
\begin{figure}[tb]
\centering
\includegraphics[height=5.7cm, width=12.00cm]{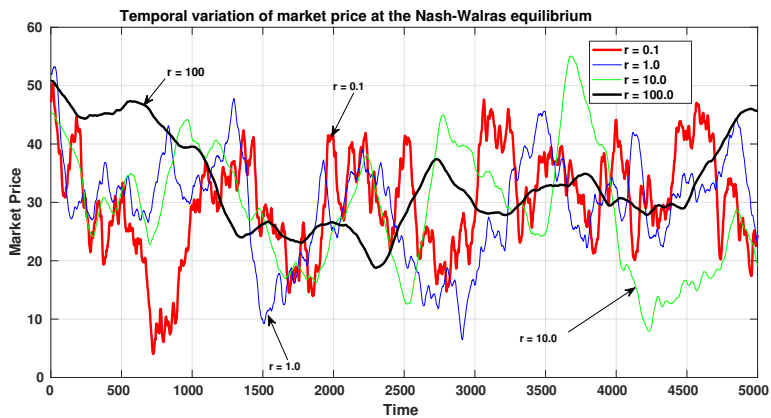} 
 \caption{State dynamics with varying control penalty coefficient $r.$ As the control penalty increases, the equilibrium price process $\left\lbrace \alpha_t,t\geq0\right\rbrace$ becomes progressively less volatile. Of particular interest are the two extremes: $r=0.1$ ({\color{red}red curve}) and $r=100$ (black curve), with $J^{\mathcal{N}}_{sp,r}= 9.552\times10^7\text{ and }1.468\times10^8$ resp.}
 \label{fig:2MarketSamplePathsWithControlPenalty}
\end{figure}

 \subsection{Market with Price-taking Prosumers}
\begin{figure*}[hbt]
\centering
\begin{minipage}[b]{.4\textwidth}
\hspace{-1.50cm}
\includegraphics[height=5.750cm, width=9.0cm]{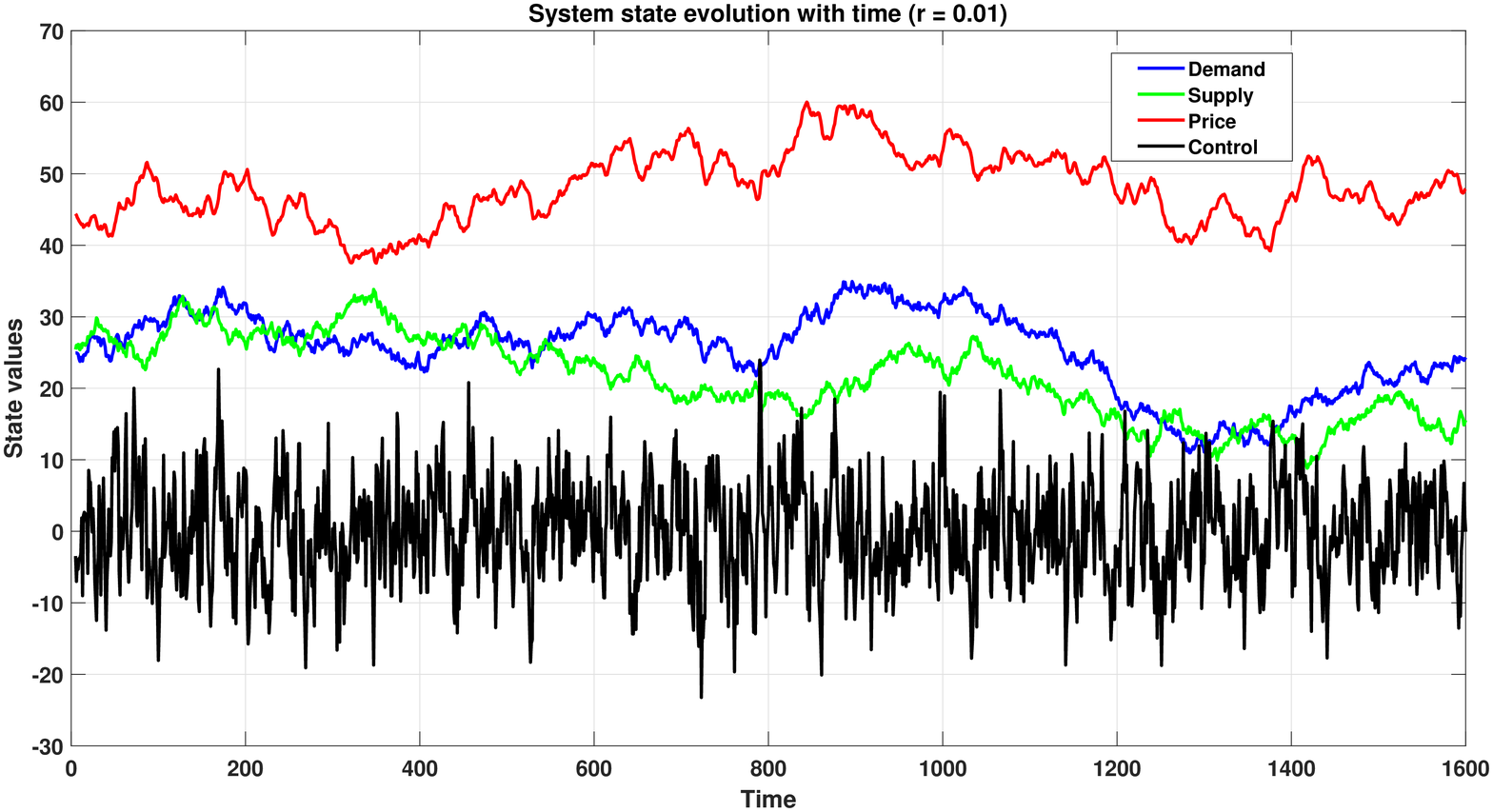} 
\caption{$r=0.01$}
\label{fig:marketDynamicsCtPenalty001}
\end{minipage}\qquad
\begin{minipage}[b]{.4\textwidth}
\includegraphics[height=5.750cm, width=9.0cm]{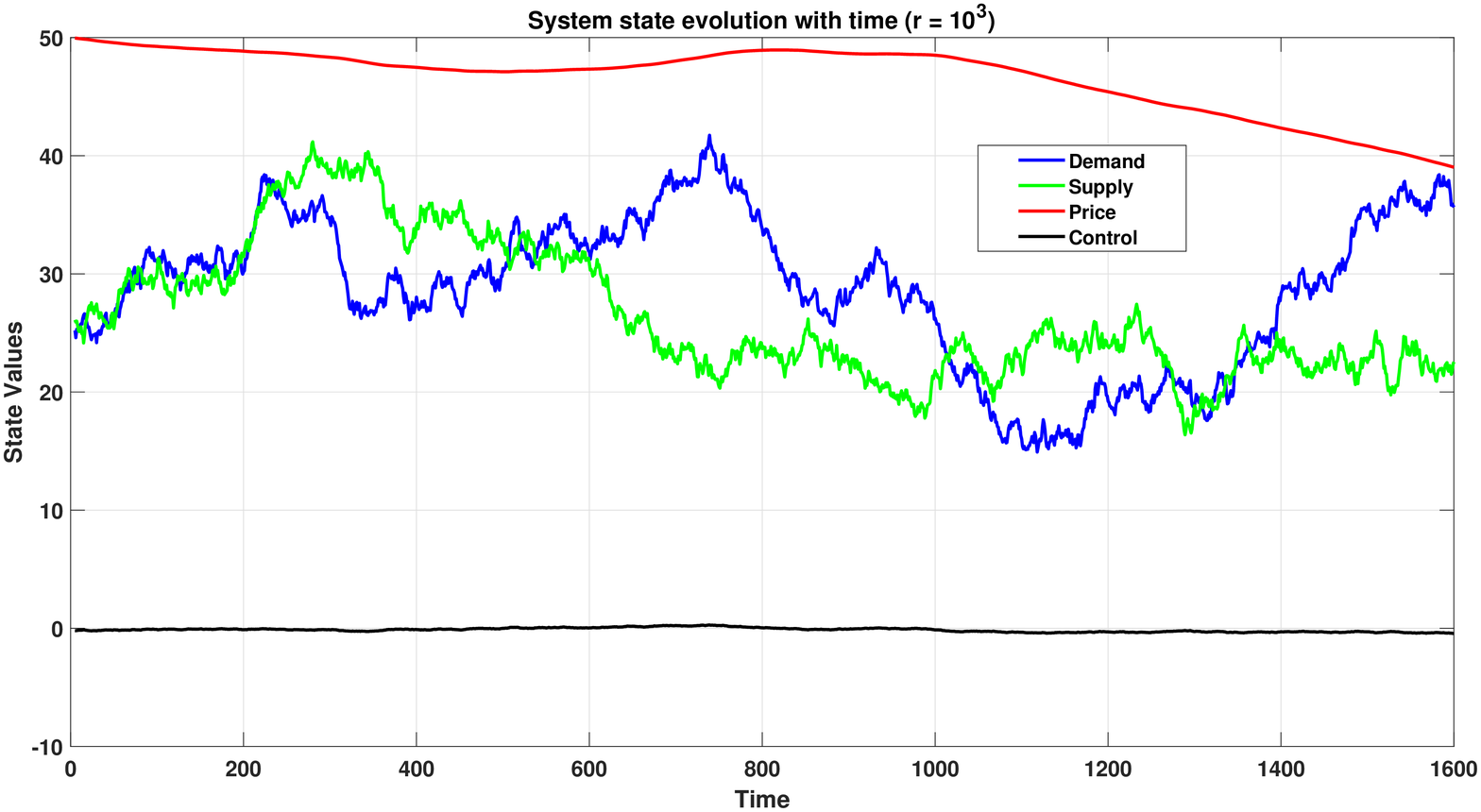}
\caption{$r=10^3$}
\label{fig:marketDynamicsCtPenalty1000}
\end{minipage}
\caption{State dynamics with varying control penalty coefficient $r.$ In Fig.~\ref{fig:marketDynamicsCtPenalty001}, the control penalty is small and variations in the price process $\left\lbrace p_t,t\geq0\right\rbrace$ and control $\left\lbrace u_t,t\geq0\right\rbrace$ are large, while in Fig.~\ref{fig:marketDynamicsCtPenalty1000}, since control fluctuations are being heavily penalized, price fluctuates less.}\label{fig:2MarketSamplePathsWithControlPenalty}
\end{figure*}

Next, we move on to a more concrete example of a market, where the state $\x_t=[d_t,s_t,p_t]$ evolves in $\reals^3$. Fig.~\ref{fig:2MarketSamplePathsWithControlPenalty} shows the effect of the control penalty coefficient $r$ in \eqref{eqn:quadraticCostWithControlPenalty}. Two sample paths of the state of the market $\x_t$ are shown with Fig.~\ref{fig:marketDynamicsCtPenalty001} demonstrating the effects of light control penalty ($r=0.01$) and Fig.~\ref{fig:marketDynamicsCtPenalty1000}, the effects of heavy control penalty ($r=10^3$). In both cases, we begin with the same initial condition $\x_0 = [25,25,50]$, but owing to the effect of the control penalty, the sample paths differ significantly in their evolution. When $r$ is small, since large values of price control $u_t$ are not penalized much, the price signal can be varied much more freely with the result that the market is more \enquote{stable,} in the sense that demand and supply do not fluctuate wildly over time.

The opposite is true in Fig.~\ref{fig:marketDynamicsCtPenalty1000}. When $r=10^3,$ the regulatory authority is highly constrained in the control that it can exercise upon the price signal and, consequently, the market is also less stable with demand and supply varying much more with time. For the sample paths in Fig.~\ref{fig:marketDynamicsCtPenalty1000}, the price volatility $\mathcal{V}$ (defined in  Eqn.~\ref{eqn:defnOfPolicyVolatility}) when $r=0.01$ is $862.16,$ which is significantly larger than the volatility level of $0.07$ obtained with $r=10^3$. Thus, the experiment clearly shows that imposing a heavier control penalty (1) reduces price fluctuations, but (2) increases demand and supply volatility, which is essentially what Thm.~\ref{thm:CostWithControlPenaltyConcaveInR} asserts. 

\begin{figure}[htb]
\centering
\includegraphics[height=5.70cm, width=12.00cm]{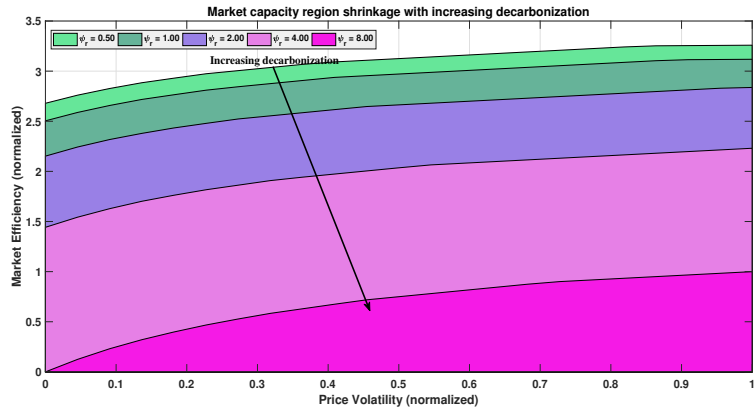}
\caption{The capacity region for the market $(A,\b,F_n,Q,r,\gamma)$ shrinks as renewables penetrate the market. We see that as $\psi_r$, the renewable source contribution to the market increases from $0.5$ to $8.0,$ the capacity region shrinks slowly at first, but rapidly thereafter, until around $\psi_r=8.0,$ barely any of the original efficiency (i.e., in $\mathcal{C}^{(0.5)}$) is achievable.
The regions have been normalized such that $\mathcal{C}^{(8.0)}$ peaks at 1.}
\label{fig:capacityRegionShrinksWithIncreasingRenewables}
\end{figure}

In Sec.~\ref{sec:renewablesAndTheVolatilityCliff} we described the deleterious effects that unchecked penetration of renewable energy sources can have on market efficiency and price volatility. Recall that in Prop.~\ref{prop:moreRenewablesMorePriceVolatility}, we already argued why increasing renewable supply penetration (modeled by increasing $\psi_r$) results in higher price volatility. Here, we go further and present a simulation supporting a stronger claim. Let $\mathcal{C}^{\left(\psi_r\right)}$ denote the capacity region associated with renewables penetration level of $\psi_r = \mathbb{E}w_t^2,$ where the latter is defined in \eqref{eqn:renewableSupplyStateEvolution}. Fig.~\ref{fig:capacityRegionShrinksWithIncreasingRenewables} illustrates precisely how quickly the achievable efficiency is lost when renewable sources increase in the current market structure. As seen in the figure, as $\psi_r$, the contribution of renewable sources, increases, the associated capacity regions shrink, with the loss in capacity becoming much more pronounced with increasing $\psi_r.$ In the Supplementary Material, aided by a more realistic market model, we show a starker version of this result (see Sec.~\ref{sec:supplementaryMaterial}). 
These observations add to the increasing groundswell in support of fundamental restructuring of electricity markets to survive the imminent influx of green power generation technologies \cite{roberts19clean-energy-technologies-threaten-adapt,joskow19challenges-wholesale-intermittent-renewable-at-scale}.
\section{Concluding Remarks and Future Work}\label{sec:conclusionsAndFutureWork}
In this paper we presented an analysis of the time slotted Linear Quadratic Regulator with the aim to examine the volatility behavior of its control policies. We first showed that regulator efficiency and control volatility are actually inextricably related in a natural tradeoff, where penalizing one is always to the detriment of the other. We also proved that associated with every LQR, is a capacity region that dictates all (volatility, efficiency) tuples that can possibly be achieved.

We then applied these results to the analysis of real-time deregulated electricity markets, with the aim to uncover the reasons for frequently observed sharp fluctuations in market prices. Therein, we showed a similar tradeoff between social welfare and the volatility of electricity prices. Furthermore, our analysis also showed that no admissible pricing mechanism can achieve simultaneously high social welfare and low volatility, which means that efficiently operating such a market will naturally and invariably lead to highly volatile electricity prices. We also showed that the entry of renewable sources into the market can harm price stability. Going into a renewables-rich future, we therefore need a firm theoretical scaffolding to help construct a new grid to cope with all these new phenomena. 

Moving forward, an interesting question that we will attempt to answer is whether this volatility-efficiency tradeoff is unique to LQRs or do other controlled dynamical systems also exhibit such behavior. Secondly, in the context of spot pricing, we studied consumers and producers who were \emph{fully aware} of the market dynamics, i.e., the system parameters were known to all participants and the RTO/ISO during operation. We would like to investigate if these results hold under  \emph{partial observability} of the system. Indeed, while each prosumer can be expected to know its own system evolution fully and instantaneously, knowledge of the dynamics of the other participants could be construed as a strong assumption. We hope to remove this in future work.

One technique proposed in \cite{kristov-etal16tale-two-visions-decentralized-transactive}  to handle increasing renewable supply penetration calls for a departure from what they term the \enquote{The Grand Central Optimization} (GCO) paradigm to a \enquote{Layered Decentralized Optimization} (LDO) paradigm. The former, GCO, refers to the existing electricity market structure where a centralized authority (RTO/ISO) is given all system information and takes all pricing and demand-supply balance decisions. It is argued in \cite{kristov-etal16tale-two-visions-decentralized-transactive} that as the number of intermittent renewable sources (especially on the demand side) increases, gathering information about the entire state space in real time and optimizing over such a large state space might become infeasible. Therefore, a decentralized \emph{tiered} architecture (LDO) with hierarchical decision-making entities is designed as an alternative. This new market needs to be analyzed in greater detail. Further, Dual Pricing is another option that as been explored in the literature, for example in \cite{kizilkalemannor11regulation-double-price-mechanisms}, as a potential method to reduce price volatility.

  \bibliographystyle{alpha}
  \bibliography{lqr_volatility_techreport}
\newpage
\section{Appendix}
\subsection{Glossary of Notation}\label{sec:appendixGlossaryNotation}
\begin{itemize}
    \item $\x_t$: the state of the discrete time market model.
    \item $\xc_t:$ the state of Consumer~$i$, $1\leq i\leq N_c.$
    \item $\xp_t:$ the state of Supplier~$j$, $1\leq j\leq N_p.$
    \item $\mathsf{N} = \mathbb{N}\cup\{0\} = \{0,1,2,\cdots\}$ the set of natural numbers including $0.$
    \item $A$: the gain matrix in the state evolution equation \eqref{eqn:LinearSystemWithConstForcingFunction} relating current state to the state in the next time step.
    \item $A^{c_i}:$ the gain matrix in the state evolution equation of Consumer~$i$ (Eqn.~\ref{eqn:consumerStateEvolution}).
    \item $A^{p_j}$: the gain matrix in the state evolution equation of Supplier~$j$ (Eqn.~\ref{eqn:producerStateEvolution}).
    \item $\b$: the coefficient of the price control input in \eqref{eqn:LinearSystemWithConstForcingFunction}.
    \item $\mathbf{n}$: the state disturbance process.
    \item $F_n$:the probability distribution function of $\mathbf{n}.$
    \item $\Psi_n$: the covariance matrix of $\mathbf{n}.$
    \item $\gamma$: the discount factor in the Markov decision process formulation \eqref{eqn:gammaDiscountedCostCENTRALIZEDLinearQuadraticSystemWithControlPenalty}, $\gamma\in(0,1).$
    \item $r$: control volatility penalty.
    \item $(A,\b,F_n,Q,r,\gamma)$: an LQR is fully determined by this tuple.
    \item $\Pi$: the set of all nonanticipatory policies.
    \item $F$: the set of all Markov policies. 
    \item $D$: the set of all deterministic Markov policies.
    \item $\mathcal{A}$: the set of all admissible policies. Admissibility is defined towards the end of Sec.~\ref{sec:lqrModelPreliminaries}.
    \item $\mathbb{P}^{\pi}_{\x}$: the probability measure induced by policy $\pi$ on the space of all sample paths, when the initial state of the system is $\x$.
    \item $\mathbb{E}^{\pi}_{\x}$: the expectation operator associated with $\mathbb{P}^{\pi}_{\x}$.
    \item $\mathcal{P}(\mathbb{R})$: the set of all probability distributions on $\mathbb{R}.$
    \item $J^*_{r}$: the optimal cost functional for a given value of $r.$
    \item $\mathcal{V}_\x(\pi)$: the volatility of the price process under policy $\pi$ when the initial state of the system is $\x$ defined in \eqref{eqn:defnOfPolicyVolatility}
    \item $\mathcal{E}_\x(\pi)$: the efficiency achieved by the price process under policy $\pi$ when the initial state of the system is $\x$, defined in \eqref{eqn:defnOfPolicyEfficiency}.
    \item $\mathcal{S} = \mathbb{R}^3\times(0,\infty)$
    \item $T$: the Bellman operator defined in \eqref{eqn:BellmanOperator}.
    \item $J_{sp,r}$: the state penalizing cost defined in \eqref{eqn:gammaDiscountedCostCENTRALIZEDLinearQuadraticSystemNoControlPenalty}.
    \item $\mathcal{C}$: the capacity region of the market $(A,\b,F_n,Q,r,\gamma)$.
    \item PQ graph: Price-Quantity graph
    \item $N_c$: number of consumers in the market.
    \item $N_p$: number of suppliers in the market.
    \item $N_{\mathcal{M}}=3N_c+2N_p$:
\end{itemize}
\subsection{Proof of Lem.~\ref{lem:VRhoIsACompleteMetricSpace}}\label{app:ProofOflem:VRhoIsACompleteMetricSpace}
First note that convergence in $\parallel \cdot \parallel$ implies pointwise convergence. Let $v$ be a limit point of $\mathcal{V}$. Therefore, there exists a sequence $v_k,k\geq0$ in $\mathcal{V}$ such that $\lim_{k\rightarrow\infty}\|v-v_k\|=0$. But this implies  that $\|v\|\leq\|v-v_k\|+\|v_k\|<\infty.$ Hence, $v\in\mathcal{V}.$ Moreover, $v\in\mathcal{V}\Rightarrow T v\in\mathcal{V}$, since
\begin{eqnarray}
    (Tv)(\x,r) &=& \x^TQ\x + \inf_{u\in\mathbb{R}} r~u^2 + \gamma \mathbb{E}_{\mathbf{n}}v(A\x+\b u+\mathbf{n},r)\nonumber\\
              &\leq& \x^TQ\x + \inf_{u\in\mathbb{R}} r~u^2 \nonumber\\ 
              && + \gamma \mathbb{E}_{\mathbf{n}}\|(A\x+\b u+\mathbf{n})\|^2+r^2+1\label{eqn:TvIsQuadraticInY}
\end{eqnarray}
The RHS of \eqref{eqn:TvIsQuadraticInY} is quadratic in $(\x,r)$ and, hence, $\|Tv\|<\infty$. 
\qed

\subsection{Proof of Lem.~\ref{lem:SubspaceHClosedUnderT}}\label{app:ProofOflem:SubspaceHClosedUnderT}
The closure of $\mathcal{H}$, i.e., that it contains all its limit points, follows immediately from the observation that convergence in $\parallel\parallel$ always implies pointwise convergence. Moving on to the next claim in the Lemma, observe that $\forall v\in\mathcal{H}$ and $(\x,r)\in\mathcal{S},$
\begin{eqnarray}
(Tv)(\x,r) = \x^TQ\x + \inf_{u\in\mathbb{R}} \underbrace{r~u^2}_{\text{linear in }r} + \gamma \underbrace{\mathbb{E}_{\mathbf{n}}v(A\x+\b u+\mathbf{n},r)}_{\text{concave in }r}.\nonumber   
\end{eqnarray}
Since $Tv$ is the pointwise infimum of a collection of concave functions, it is itself concave in $r.$ That $Tv$ is non decreasing whenever $v$ is, follows mutatis mutandis.
\qed
\subsection{Proof of Lem.~\ref{lem:zeroFunctionConvergesToJStar}}\label{app:ProofOflem:zeroFunctionConvergesToJStar}
It is easy to see that, starting with $v_0,$ the all zero function\footnote{Meaning $v_0(\x,r)=0,~\forall(\x,r)\in\mathcal{S}$.}, for $\forall t\geq0,$ and $\forall (\x,r)\in\mathcal{S},$
\begin{eqnarray}
  T^{(t+1)}v_0(\x,r) &=& \x^TK_t\x + \sum_{m=0}^{t-1}\gamma^{k-m}\mathbb{E}\mathbf{n}^T K_m\mathbf{n},\nonumber
\end{eqnarray}
where, the matrices $K_t$ are defined by 
\begin{eqnarray}
    K_0 &=& Q \nonumber\\
    K_{t+1} &=& Q + A^T \left[\gamma K_t - \frac{1}{\gamma\b^t K_t \b+r}\gamma^2 K_t \b\b^T K\right] A,~t\geq0.\nonumber
\end{eqnarray}
Also recall from Eqn.~\eqref{eqn:optimalCostSystemWithControlPenalty} that 
\begin{equation}
    J^*_r(\x) = \x^TK\x + \frac{\gamma}{1-\gamma}\mathbb{E}\mathbf{n}^TK\mathbf{n},~\forall(\x,r)\in\mathcal{S}.
\end{equation}
Therefore, 
\begin{eqnarray}
    \bigg|T^{(t+1)}v_0(\x,r) - J^*_r(\x)\bigg| &=& \bigg|\x^T(K_t-K)\x + \sum_{m=0}^{k-1}\gamma^{k-m}\mathbb{E}\mathbf{n}^T K_m\mathbf{n} \nonumber\\  
    && - \frac{\gamma}{1-\gamma}\mathbf{n}^T K_m\mathbf{n} \bigg| \nonumber\\
    &\leq& \underbrace{\bigg|\x^T(K_t-K)\x \bigg|}_{T1} \nonumber\\  
    && + \underbrace{\bigg|\sum_{m=0}^{k-1}\gamma^{k-m}\mathbb{E}\mathbf{n}^T K_m\mathbf{n} - \frac{\gamma}{1-\gamma}\mathbf{n}^T K_m\mathbf{n}}_{T2} \bigg| \nonumber.
\end{eqnarray}

Under observability and controllability, $K_t$ converges pointwise to $K$ \cite[Prop.~4.1]{bertsekas95dynamic-programming-optimal-control-volume1}. So, coming to term $T1$ above, we see that 
\begin{eqnarray}
  \lim_{t\rightarrow\infty} \sup_{(\x,r)\in\mathcal{S}}\frac{|\x^T(K_t-K)\x |}{\left(\|(\x,r)\|_2\vee1\right)}\leq \lim_{t\rightarrow\infty} \parallel K_t-K \parallel_{F} = 0,
\end{eqnarray}
where $\parallel A\parallel_F$ is the Frobenius norm of matrix A. Similarly, $T2$ 
vanishes as $t\rightarrow\infty$ due to the pointwise convergence of $K_t$ to $K.$
\qed

\subsection{State penalizing cost}\label{app:ProblemWithStatePenalizingCost}
\begin{prop}\label{prop:discountedCostApplyingUStarWithoutControlPenalty}
The $\gamma$-discounted cost of applying the stationary policy $\pi^*=[u^*_0,u^*_1,\cdots]$ where 
  $u^*_t(\x) = -\frac{1}{\b^t K \b+r} \b^t K A \x,$
to the system in Eqn.~\eqref{eqn:LinearSystemWithConstForcingFunction} with single-stage cost $g(\x_t,u_t) = x_t^TQx_t,t\geq0,\x\in\reals^d$ is given by \eqref{eqn:riccatiEqnForStatePenalizingCost}, i.e.,
\begin{eqnarray}
  J_{sp,r}(\x_0) &=& \x^T_0S\x_0 + \frac{\gamma}{1-\gamma}\mathbb{E}\mathbf{n}^TS\mathbf{n},
\label{eqn:appendixGammaDiscountedNoControlPenaltyInTermsOfS}
\end{eqnarray}
where $S$ satisfies
\begin{eqnarray}
S &=& Q +  \gamma A^T \left[ I - \frac{1}{\b^T K \b+r} K\b\b^T \right] S \left[ I - \frac{1}{\b^T K \b+r} \b\b^T K\right] A.\nonumber\\
\label{eqn:appendixRiccatiEqnForStatePenalizingCost}
\end{eqnarray}
\end{prop}
\begin{pf}
The function $J_{sp,r}(\cdot)$ is, by definition, the $\gamma$-discounted cost of applying the stationary control $u^*_t(\x) = -\frac{1}{\b^t K \b+r} \b^t K A \x$ to the system in \eqref{eqn:LinearSystemWithConstForcingFunction}. 
Computing $J_{sp,r}$, therefore, can be cast as applying $\pi^*$ to a $\gamma$-discounted  controlled Markov process, specified by 
\begin{itemize}
    \item State space: $\mathcal{S}=\reals^d\times(0,\infty)$
    \item Action space: $\mathsf{A} = \mathbb{R}.$ Note that $U(\x,r) = \mathbb{R}~\forall(\x,r)\in\mathcal{S}.$
    \item Transition probability kernel: Let $\mathcal{K}=\{(\mathbf{y},u)\in\mathcal{S}\times\mathsf{A}\mid u\in U(\mathbf{y})\}$ be the set of all admissible atate-action pairs, and let $\mathcal{B}(\mathcal{S})$ be the Borel sigma algebra on $\mathcal{S}$. Then, the transition probability kernel $T:\mathcal{K}\rightarrow[0,1]$ is a stochastic kernel on $\mathcal{S}$ specified for every $\left((\x,r),u\right)\in\mathcal{K}$ by 
    \begin{equation*}
        T(B\mid\left((\x,r),u\right)) = \int_{(\mathbf{y},s)\in B}\mathbb{I}_{\{s=r\}} dF_n\left(\mathbf{y}-A\x-\b u\right),~\forall B\in\mathcal{B}(\mathcal{S}),
    \end{equation*}
    where $F_n$ is the distribution of the state disturbance.
    \item The Single Stage Cost is a measurable function $g: \mathcal{K}\rightarrow\mathbb{R}_+$, defined by $g((\x,r),u) = \x^TQ\x~\forall(\x,r)\in\mathcal{S}.$
    \item Discount factor = $\gamma.$
\end{itemize}
The cost of applying $\pi^*$ to this MDP is the state penalizing cost $J_{sp,r}$. Since $\pi^*$ is a stationary policy, we invoke the following from \cite[Cor.~3.1.1.1]{bertsekas11dynamic-programming-optimal-control-volume2}.
\begin{prop}[\color{blue}\cite{bertsekas11dynamic-programming-optimal-control-volume2}]
Let $\mu$ be a stationary policy for the above MDP. Suppose the single stage cost $g(\x,u)\geq0,~\forall(\x,u)\in\mathcal{S}\times\mathbb{R}$. Then the cost functional $J_{\mu}()$ associated with $\mu$ satisfies 
\begin{equation}
    J_{\mu}(\x,r) = \x^TQ\x + \gamma \mathbb{E}J_{\mu}\left(A\x+\b \mu(\x)+\mathbf{n},r\right)\nonumber
\end{equation}
\end{prop}

This shows that under $\pi^*,$ $J_{sp,r}$ satisfies 
\begin{equation}
    J_{sp,r}(\x) = \x^TQ\x + \gamma \mathbb{E}_{\mathbf{n}}J_{sp,r}\left(A\x+\b u^*(\x)+\mathbf{n}\right).
    \label{eqn:bellmanForStatePenalizingCost}
\end{equation}
A substitution argument now shows that $J_{sp,r}(\x) = \x^TS\x + \frac{\gamma}{1-\gamma}\mathbb{E}\mathbf{n}^TS\mathbf{n}$, where the matrix $S$ satisfies \eqref{eqn:appendixRiccatiEqnForStatePenalizingCost}.

\end{pf}

\subsection{Proof of Prop.~\ref{prop:existenceOfSolutionToRiccatiRecursionWithoutControlCost}}\label{app:ProofOfPropExistSolRiccatiWithoutCtrlCost}
This proof is similar to that of Prop.~4.4.1 in \cite{bertsekas95dynamic-programming-optimal-control-volume1} and proceeds in several steps. Recall that the iteration under study is given by
\begin{eqnarray}
\hspace{-1.00cm}
  S_{t+1} &=& Q +  \left(\gamma A^T \left[ I - \gamma\frac{1}{\gamma\b^T K \b+r} K\b\b^T \right]\right.\nonumber\\
          && \left. S_{t} \left[ I - \gamma\frac{1}{\gamma\b^T K \b+r} \b\b^T K\right] A\right),~t\geq0,
  \label{app:eqn:recursion4riccatiEqnForStPenCost}
\end{eqnarray}
with $S_0$ set to any SPSD matrix. We first prove convergence starting with $S_0=0_{3\times3}$, the latter being the all zeros matrix in $\mathbb{R}^{3\times3},$ and then show convergence with an arbitrary initial matrix $S_0.$ Since this means we'll be dealing with multiple initial conditions for the above iteration, we denote the $t^{th}$ iterate by $S_t(S_0)$ to make explicit the initial condition. So, for example, the iterate beginning with the all zeros matrix will be denoted by $S_t(0_{3\times3}).$

Before we begin, notice that under the controllability and observability assumptions, the closed loop (noiseless) system given by 
\begin{eqnarray}\label{app:eqn:linearNoiselessSystem}
  \x_{t+1} = A\x_t + \b u_t,~t\geq0
\end{eqnarray}
is {\color{blue}stable} \cite[Prop.~4.4.1 (b)]{bertsekas95dynamic-programming-optimal-control-volume1} in the sense that when $$u_t = -\gamma\frac{1}{\gamma\b^t K \b+r} \b^t K A \x_t = G\x_t,$$ 
the matrix 
\begin{equation}
D = G + A = \left[ I - \gamma\frac{1}{\gamma\b^T K \b+r} \b\b^T K\right] A    
\end{equation}
has all three eigenvalues inside the unit circle. Denote the Jordan decomposition \cite[Chap.~3]{horn-johnson12matrix-analysis} of $D$ by $D=P^{-1}\Lambda_D P$, and note that the previous remark implies that $\lim_{k\rightarrow\infty}D^k = \lim_{k\rightarrow\infty}\Lambda_D^k = 0_{3\times3}$ (again, this is entrywise convergence and hence, in Frobenius norm).\\
    \textbf{Step 1:} Let $S_0=0_{3\times3}.$ Then, $S_t=\sum_{k=0}^{t-1}\gamma^k(D^T)^kQD^k,~t\geq1.$ Now consider, for the time being, the problem of applying the $u_t=G\x_t,t\geq0$ to the system in \eqref{app:eqn:linearNoiselessSystem} for exactly $t$ time steps\footnote{This means we're now working with a \emph{finite} horizon version of the problem, with the horizon being $t$ time steps.}, with single stage cost 
    \begin{eqnarray}
    g(\x_k,u_k) &=& \x_k^TQ\x_k,~0\leq k\leq t-1,\nonumber\\ 
    g(\x_t) &=& \x_t^TS_0\x_t\nonumber,
    \end{eqnarray}
     and $J^{\pi}_t(\x_0)=\sum_{k=0}^{t-1}\gamma^kg(\x_k,u_k)+\gamma^t\x_tS_0\x_t$. Then we see that under our policy, $J^{\pi}_t(\x_0)=\x_0S_t(0_{3\times3})\x_0$, 
    which means that $\x^TS_t(0_{3\times3})\x\leq\x^TS_{t+1}(0_{3\times3})\x,~t\geq0,~\x\in\mathbb{R}^3.$ So $\x^TS_t(0_{3\times3})\x$ is a nondecreasing sequence.
    
    Under our control policy, 
    \begin{eqnarray}
      \x_{t+1} &=& D\x_t = P^{-1}\Lambda_D P\x_t,~t\geq0,\nonumber\\
      \Rightarrow \mathbf{y}_{t+1} &=& \Lambda_D\mathbf{y}_t,~t\geq0,\nonumber\\
      \Rightarrow \mathbf{y}_{t} &=& \Lambda_D^t\mathbf{y}_0,~t\geq1,\nonumber
    \end{eqnarray}
    where $\mathbf{y}_t = P\x_t.$ So the $t$ stage cost can be rewritten as 
    \begin{eqnarray}
    J^{\pi}_t(\x_0) &=& \x_0^TS_t(0_{3\times3})\x_0\nonumber\\
                    &=& \sum_{k=0}^{t-1}\gamma^k\x_0^T(D^T)^kQD^k\x_0\nonumber\\
                    &=& \sum_{k=0}^{t-1}\gamma^k\mathbf{y}_0^T(\Lambda_D^T)^kP^{-1}QP\Lambda_D^k\mathbf{y}_0\nonumber\\
                    &\stackrel{(\ast)}{\leq}& \lambda_{max}(P^{-1}QP)\parallel\mathbf{y}_0\parallel^2_2\sum_{k=0}^{t-1}\gamma^k|\lambda_{max}(\Lambda_D)|^k\nonumber.
    \end{eqnarray}
    In $(\ast)$, $\lambda_{max}(P^{-1}QP)$ and $\lambda_{max}(\Lambda_D)$ are the largest eigenvalues (in modulus) of $P^{-1}QP$ and $\Lambda_D$ respectively. Since all the eigenvalues of $D$ are within the unit circle, the above sum converges as $t\rightarrow\infty.$ 
    
    So we now have $\left\lbrace\x_tS_t(0_{3\times3})\x_t\right\rbrace_{t=0}^\infty$ being a nondecreasing sequence bounded from above, which means that it converges (which is also what the above inequalities show). Hence, beginning with the all zeros matrix, the iteration \eqref{app:eqn:recursion4riccatiEqnForStPenCost} converges to some limit, say $S$, that satisfies \eqref{eqn:riccatiEqnForStatePenalizingCost}. Using different values for the vector $\x_0$, it can be shown that this convergence is in the Frobenius norm.
    
    \textbf{Step 2:} Let $S_0$ be any arbitrary symmetric positive semidefinite matrix. Going back to the $t$ step finite horizon formulation in Step 1, we see that, modifying the final stage cost to $g(\x_t) = \x_t^TS_0\x_t$, 
    \begin{eqnarray}
      \x^TS_t(S_0)\x = \underbrace{\x \gamma^t(D^t)^TS_0D^t}_{T1} + \underbrace{\sum_{k=0}^{t-1}\gamma^k\x(D^T)^kQD^k\x}_{T2}.\nonumber
    \end{eqnarray}
    In the above equation, $T1\rightarrow0$ and we have already shown that $T2$ converges. This completes the proof of convergence of the iterates in \eqref{app:eqn:recursion4riccatiEqnForStPenCost}.
Finally, to show uniqueness, assume the contrary, i.e., there exists another $S'$ that is a fixed point of Eqn.~\eqref{app:eqn:recursion4riccatiEqnForStPenCost}. Then, beginning \eqref{app:eqn:recursion4riccatiEqnForStPenCost} with $S_0 = S',$ we see that $S_t(S')\rightarrow S,$ which means $S' = S.$ Hence, the solution to \eqref{eqn:riccatiEqnForStatePenalizingCost} is unique.
\qed

\subsection{Proof of Thm.~\ref{thm:CostWithOutControlPenaltyConcaveInR}}\label{app:ProofOfCostWithoutControlPenaltyConcave}
We first define the norm $\parallel\parallel$ as in Eqn.~\eqref{eqn:supNormOnFunctionSpaces}, the space $\mathcal{V}$ and the metric $\rho$ exactly as in the proof of Thm.~\ref{thm:CostWithControlPenaltyConcaveInR}. We also note that $\langle\mathcal{V},\rho\rangle$ is a complete metric space and the set $\mathcal{H} := \{v\in\mathcal{V}: v\text{ is concave nondecreasing}\}$ is closed in $\mathcal{V}$. Next, recall that $D = \left[ I - \frac{1}{\b^T K \b+r} \b\b^T K\right] A$ and define the operator $T_{\pi}:\mathcal{J}\rightarrow\mathcal{J}$ by
\begin{eqnarray}
  T_\pi v(\x,r) = \x^TQ\x + \gamma v\left(D\x,r\right).
\end{eqnarray}
The fact that $v\in\mathcal{V}\Rightarrow T_\pi v\in\mathcal{V}$ follows using the ideas in the proof of Lem.~\ref{lem:VRhoIsACompleteMetricSpace} and that $v\in\mathcal{H}\Rightarrow T_\pi v\in\mathcal{H}$ from the proof of Lem.~\ref{lem:SubspaceHClosedUnderT}. 

Next, observe that the all zero cost function $v_0$ defined by $v_0(\x,r)=0,~\forall (\x,r)\in\mathcal{S}$ is in $\mathcal{H}.$ So, we now only need to show that, beginning with $v_0$, $T^kv\rightarrow J_{sp,r}.$ This will follow from the proof of Lem.~\ref{lem:zeroFunctionConvergesToJStar}, if we can show that $\parallel S_t-S\parallel_F\rightarrow0,$ but that is precisely what Prop.~\ref{prop:existenceOfSolutionToRiccatiRecursionWithoutControlCost} states. Hence, $J_{sp,r}\in\mathcal{H}$ which concludes the proof of the theorem.
\qed
\subsection{Proof of Thm.~\ref{thm:CapacityRegionOfMarket}}\label{app:ProofOfThmCapacityRegionOfMarket}
We have proved that $\mathcal{E}^*_\x$ is concave nondecreasing in volatility. Basic convex analysis shows that the epigraph of a convex function is always convex \cite{bertsekas-nedich03convex-analysis-optimization} which, therefore is true of the the region bounded \emph{below} this \emph{concave} curve. Since at optimality the volatility constraint is active, the boundary of $\mathcal{C}$ is achieved by the policy that attains the optimum in \eqref{eqn:optimalCostOfConstrainedMDP} which, by Thm~4.2 in \cite{altman99constrained-markov-decision-processes}, exists. This means that, $\mathcal{C},$ the set of $(\mathcal{E},\mathcal{V})$ pairs that can be achieved, is a subset of the subset of $\mathbb{R}^2$ bounded by this curve. Therefore, to complete the proof, we need to show that every point in the region below this curve is achievable.

Recall that at optimality, we require that 
\begin{equation}
    \x_0\frac{dK_\lambda}{d\lambda}\x_0 + \frac{\gamma}{1-\gamma}\mathbb{E}\mathbf{n}^T \frac{dK_\lambda}{d\lambda}\mathbf{n} = \alpha,
    \label{eqn:atOptimalityDerivativeWrtLambdaEqualsVolatilityAlpha}
\end{equation}
and that $\x\frac{dK_\lambda}{d\lambda}\x$ is decreasing in $\lambda.$ This means that as volatility $\alpha\downarrow0$, to satisfy \eqref{eqn:atOptimalityDerivativeWrtLambdaEqualsVolatilityAlpha} above, $\lambda\uparrow\infty,$ which means that $J^*_r\uparrow\infty\text{ and }\mathcal{E}^*_r\downarrow-\infty.$ This means that the capacity region is convex, but {\color{blue} not compact}. Moreover, since $\mathcal{E}^*_r$ is a nonpositive, nodecreasing function of $\alpha,$ it has a limit point. Now, let $\mathcal{C}^o$ denote the interior of $\mathcal{C}.$ It is easy to see that for every point $\mathbf{z}=(v,e)\in\mathcal{C}^o,$ there exist points $\mathbf{z}_1=(v_1,e^*_1)$ and $\mathbf{z}_2=(v_2,e^*_2)$ such that $e^*_i,~i=1,2$ is the optimal efficiency for volatility $v_i$ and $\mathbf{z}=\mu\mathbf{z}_1+(1-\mu)\mathbf{z}_2$ for some $\mu\in[0,1].$ We need to show that there exists some admissible policy that achieves $(v,e)$.

Towards this end consider the policy $\pi_{\mu}$ defined as follows. Let $\pi_i,~i=1,2$ be the policy that achieves $(v_i,e^*_i).$ At time $t=0,$ the policy $\pi_\mu$ chooses $\pi_1$ with probability $\mu$ and $\pi_2$ with probability $1-\mu$ and continues with this choice for all time slots thereafter. The associated expectation operator is $\mathbb{E}^{\pi_\mu}_\x(\cdot) = \mu \mathbb{E}^{\pi_1}_\x(\cdot) + (1-\mu)\mathbb{E}^{\pi_2}_\x(\cdot)$. Such a policy obviously achieves efficiency $\mu e^*_1 + (1-\mu)e^*_2$ and volatility $\mu v_1 + (1-\mu) v_2$ (which are both finite), and is therefore also admissible. 

\subsection{Proof of Prop.~\ref{prop:moreRenewablesMorePriceVolatility}}\label{appendix:proofOfPropmoreRenewablesMorePriceVolatility}
It is easy to check that the conditions for controllability and observability do not change from the ones in Sec.~\ref{sec:lqrModelPreliminaries}, and therefore, going through the constrained MDP analysis (see Sec.~\ref{sec:CapacityRegionNoFreeLunchTheorem}) once more on this new system, the optimal cost is given by
\begin{eqnarray}
  \mathcal{L}^{(ren),*}_r(\mathbf{x}) &=& \sup_{\lambda\geq0}\left(\mathbf{z}^TK^{(r)}_{\lambda}\mathbf{z}  +  \frac{\gamma}{1-\gamma}\mathbb{E}(\mathbf{n}^{(r)})^TK^{(r)}_{\lambda}\mathbf{n}^{(r)}\right. \nonumber\\
  && \left.- \lambda\alpha\right),~\forall\x_0\in\mathbf{R}^3.
  \label{eqn:renewablesConstrMDPOPtimalCost}
\end{eqnarray}
As renewable supplies proliferate (a) the aggregate supply from these sources obviously increases and (b) supplied power becomes more volatile. The factor in Eqn.~\eqref{eqn:renewablesStateEvolution} that captures both of these effects is the renewable supply variance $\psi_r.$ As the contribution of renewables increases, so does $\psi_r.$ Returning to Eqn.~\eqref{eqn:renewablesConstrMDPOPtimalCost}, notice that the contribution of the state disturbance terms to the optimal cost is $\mathbb{E}(\mathbf{n}^{(r)})^TK^{(r)}_{\lambda}\mathbf{n}^{(r)}$ (ignoring the constant factor), which can be rewritten as $\mathbb{E}Tr\left(K^{(r)}_{\lambda}(\mathbf{n}^{(r)})^T\mathbf{n}^{(r)}\right) = Tr\left(K^{(r)}_{\lambda}\Psi^{(r)}_n\right)$, where $Tr(M)$ is the trace of the matrix $M.$ Since $\Psi^{(r)}_n$ is diagonal, it is clear that as $\psi_r$ increases, so does $Tr\left(K^{(r)}_{\lambda}\Psi^{(r)}_n\right)$, which increases the optimal cost. 

We already know from our analysis in Sec.~\ref{sec:CapacityRegionNoFreeLunchTheorem} that $\mathcal{L}^{(ren),*}_r(\mathbf{x})$ is nondecreasing in $\alpha$ and hence, when $\psi_r$ increases, the only way to maintain cost, i.e., social utility, the same is to increase $\alpha.$ But by definition, an increase in $\alpha$ automatically implies increased volatility in prices. 


\subsection{Details of Experiments}\label{app:detailsOfeExperiments}
The market, as mentioned before, is completely specified by the tuple $(A,\b,F_n,Q,r,\gamma)$. Recall that the matrix $A$ is of the form 
\begin{eqnarray*}
A =\left[ \begin{matrix}
  \beta & 0 & -\phi_1 \\
  0 & \sigma & \phi_2 \\
  0 & 0 & 1
 \end{matrix}\right],
\end{eqnarray*}
where $\beta,\sigma,\phi_1,\phi_2$ are nonnegative reals. In our experiments, we have chosen $\beta = 0.995, \sigma = 0.900, \phi_1 = 0.5, \phi_2 = 0.25$ and the discount factor $\gamma = 0.5$. The controllability matrix turns out to be
\begin{eqnarray*}
\left[ \begin{matrix}
  0 & -0.5 & -0.9975 \\
  0 & 0.25 & 0.475 \\
  1 & 1 & 1
 \end{matrix}\right],
\end{eqnarray*}
which has a rank of 3, and hence, the system is controllable. Since we have performed experiments for multiple values of the control penalty $r$, these values are specified in the description of each experiment in Sec.~\ref{sec:simulationResults}. The state disturbance is assumed to be IID Gaussian, with 0 mean and covariance matrix\footnote{Choosing a slightly larger value of variance than $1.00$ for the state noise helps illustrate the effects of small $r$ more clearly.}
\begin{eqnarray*}
 \Psi_n =\left[ \begin{matrix}
  2.00 & 0 & 0 \\
  0 & 2.00 & 0 \\
  0 & 0 & 0
 \end{matrix}\right],
\end{eqnarray*}
 so $F_n\equiv \mathsf{N}(\mathbf{0},\Psi_n).$ 
 With regards to the cost function, to ensure observability, $Q$ was designed to be symmetric positive definite with a minimum eigenvalue of $0.5$
 \begin{eqnarray*}
 Q =\left[ \begin{matrix}
  2.38 & -1.73 & -0.15 \\
  -1.73 & 2.15 & 0.16 \\
  -0.15 & 0.16 & 0.52
 \end{matrix}\right].
\end{eqnarray*}



 \newpage
 \section{Supplementary Material: The Case for a Volatility Cliff}\label{sec:supplementaryMaterial}
 
In Sections \ref{sec:renewablesAndTheVolatilityCliff} and \ref{sec:simulationResults} we studied the effect of the presence of large-scale renewable sources (such as wind farms and solar parks) on the \emph{supply} side of our deregulated market. But apart from large-scale renewable generation, the electricity market has recently also witnessed burgeoning power supply local to the demand side or customer side, such as isolated solar panels and small wind turbines. In this section, we would like to report some preliminary results of studying the effect such sources could have on the volatility of real-time electricity prices. These so-called Distributed Energy Resources or DERs  \cite{roberts19clean-energy-technologies-threaten-adapt,kristov-etal16tale-two-visions-decentralized-transactive}, are generally located \enquote{behind the meter,} have nearly zero marginal operating costs \cite{joskow19challenges-wholesale-intermittent-renewable-at-scale} and their output is largely unaffected by grid state variables such as the instantaneous price of electricity. Consequently, studying the effect of such sources will require some modification to our original system model in Eqn.~\eqref{eqn:LinearSystemWithConstForcingFunction}. 
 
 To incorporate the aforementioned features of DERs, we propose the following market model. 
 As in Sec.~\ref{sec:renewablesAndTheVolatilityCliff}, we introduce a new state variable, $y_t$, that represents DER supply and evolves as 
 \begin{equation}
     y_{t} = \sigma_{rn} P_{rn}(t) + w_t,
 \end{equation}
where $ \sigma_{rn} P_{rn}(t)$ represents the contribution of the nominal output of the DERs to the aggregate market demand, and $\left(w_t\right)_{t=0}^\infty$ is a sequence of zero-mean independent, identically distributed random variables that models random temporal variability (caused by weather conditions, etc.). In contrast to earlier notation, we denote the variance of $w_t$ by $\psi_w.$ To account for the fact that this portion of the load is being handled at the consumer end, the market demand evolution needs to be modified to
\begin{equation}
    d_{t+1} = (\beta d_t - \phi_1 p_t - y_t)^+,
    \label{supplementary:newDemandEquation}
\end{equation}
where for any $x\in\mathbb{R},$ $(x)^+=\max(x,0).$ Notice that, in the absence of the DER contribution term $y_t,$ Eqn.~\ref{supplementary:newDemandEquation} boils down to our original demand equation in \eqref{eqn:LinearSystemWithConstForcingFunction}. 

Nominal renewable generation (especially through photovoltaic sources) tends to show a step-like behavior over a period of $T=24$ hours, peaking around midday. As empirical data from the California ISO shows \cite{caiso20todays-supply}, renewable supply satisfies approximately $10\%$ of the total demand over about $14$ hours and approximately $44\%$ for during the other $10,$ and we model this by setting
\[
    P_{rn}(t) = \left\lbrace\begin{array}{lr}
        v_1, & \text{ if }t<0.3T\text{ or }t>0.7T\\
        v_2, & \text{ if }t\in[0.3T,0.7T]\\
        \end{array}
        \right. ,
        \tag{$\dagger$}
    \label{eqn:renewableSupplyAsAFunctionOfTimeOfDay}
  \]
  Further, it has been observed that apart from the current demand, the price of electricity is also dependent on load \emph{fluctuations} \cite{wei-etal14competitive-equilibrium-electricity-price-fluctuation}. We therefore modify the price evolution equation in \eqref{eqn:LinearSystemWithConstForcingFunction} to reflect this observation as follows
  \begin{equation}
      p_{t+1} = p_t + u_t + \mathcal{D}\left(d_t,d_{t+1}\right),
  \end{equation}
  where, as in \cite{wei-etal14competitive-equilibrium-electricity-price-fluctuation}, the function $\mathcal{D}(\cdot,\cdot):\mathbb{R}^2\rightarrow\mathbb{R}_+$ is assumed to be jointly convex in its arguments and monotonically increasing in the absolute value of the difference in demand, i.e., if $|x-y|>|x'-y'|$ then $\mathcal{D}(x,y)>\mathcal{D}(x',y').$ 
  
\begin{figure}[bh]
\centering
\includegraphics[height=6.0cm, width=9.5cm]{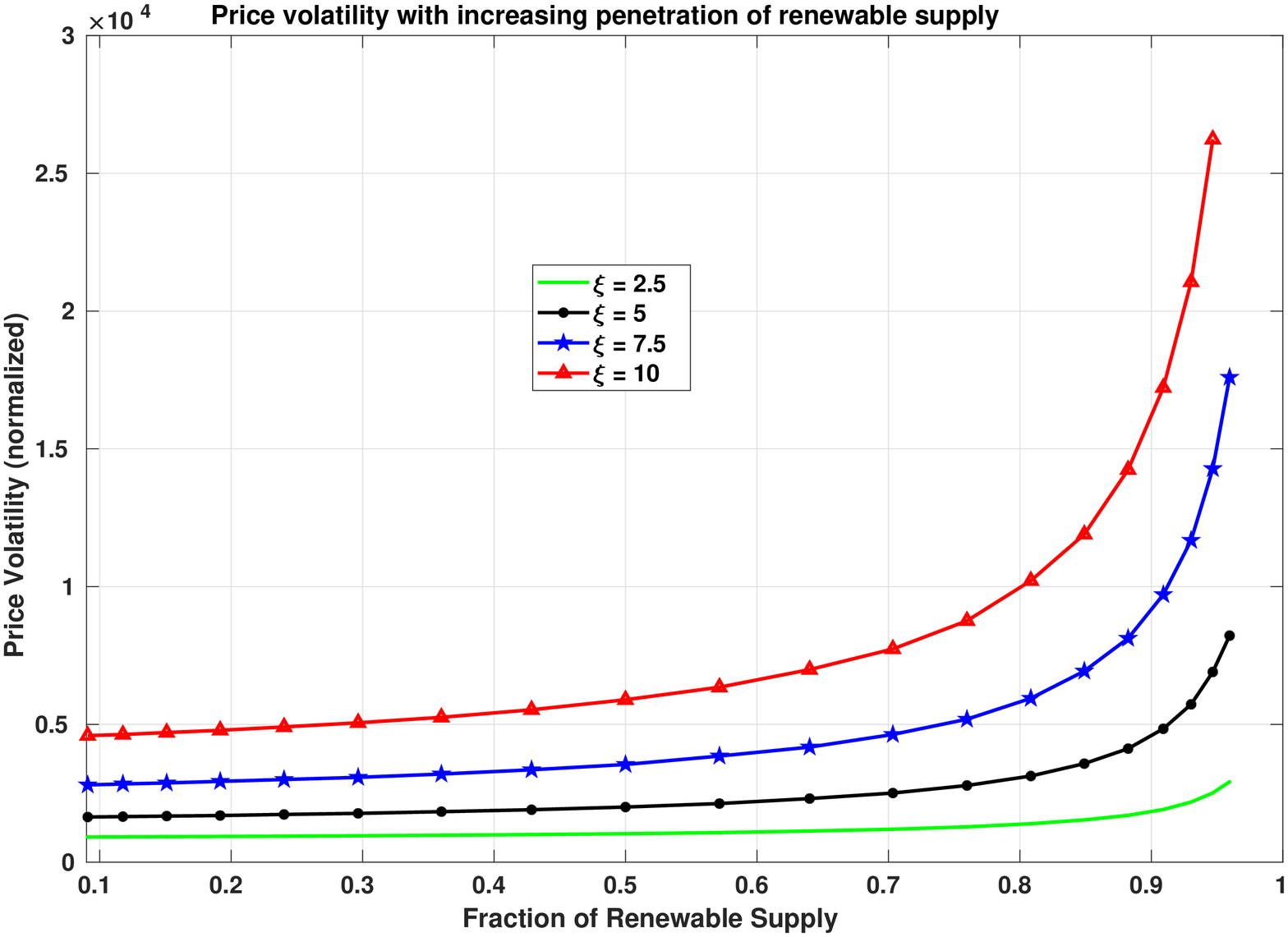}
\caption{The rapid increase of market price volatility with increasing demand-side renewable penetration. We used $\sigma_{rn} = 1.00,$ and in Eqn.\eqref{eqn:renewableSupplyAsAFunctionOfTimeOfDay}, we set $v_1 = 0.1$ and $v_2 = 0.44$. The initial condition $\x_0 = [1,1,2]^T.$ The other parameters of the market are specified in Sec.~\ref{app:detailsOfeExperiments} in the Appendix.}
\label{fig:supplementaryVolatilityIncreasesWithIncreaseInRenewableSupply}
\end{figure}

We now illustrate what happens when an RTO or an ISO, oblivious to the presence of demand-side renewable supply, decides to employ the optimal control policy discussed in Sec.~\ref{sec:lqrModelPreliminaries} on this market. In Fig.~\ref{fig:supplementaryVolatilityIncreasesWithIncreaseInRenewableSupply}, we plot the increase in price volatility with increasing renewable supply. Recall that the variance of the state disturbance for the supply process was denoted by $\psi_s,$ and let $\Delta = \frac{\psi_w}{\psi_w+\psi_s}$ represent the fraction of renewable supply in the market. Fig.~\ref{fig:supplementaryVolatilityIncreasesWithIncreaseInRenewableSupply} now shows how market price shows unbounded variability as $\Delta$ increases to $100\%$ of the market share. We choose 
\begin{eqnarray}
  \mathcal{D}(d_t,d_{t+1}) := \xi (d_t-d_{t+1})^2
\end{eqnarray}
to model price variations due to demand fluctuations. The rest of the parameters of the market are as given in Sec.~\ref{app:detailsOfeExperiments}. The figure shows that volatility increases to unacceptable levels even for moderately small values of $\xi.$ This rather precipitous increase in volatility is why we choose to term this phenomenon a \enquote{{\color{blue}volatility cliff},} since, beyond a certain level of $\Delta,$ the variability rise is high enough to force the dispatch of additional \emph{controllable} power supplies, such as coal or natural gas,to reduce $\Delta$ and hence, volatility. 

Price fluctuations in real time electricity markets have been observed and reported in, for example, the New England ISO \cite{klitgaard-etal00lowering-electricity-prices-deregulation} \emph{before} renewable proliferation became significant. But the volatility reported therein (see the section titled \enquote{Price Volatility} in \cite[Pg.~5]{klitgaard-etal00lowering-electricity-prices-deregulation}), is much smaller than that predicted by our model. This suggests that renewable supplies on the demand side, especially intermittent photovoltaic and wind sources, could be harmful to deregulated power markets if adopted without due consideration. 

\end{document}